\documentclass[%
 reprint,
 amsmath,amssymb,
 aps,
]{revtex4-2}

\usepackage{graphicx,xspace} 

\usepackage{natbib} 
\newcommand{\angmom}{\mathcal{L}}

\newcommand{\changaMM}{\texttt{{MANGA}}\xspace}
\newcommand{\changa}{{\texttt{ChaNGa}}\xspace}
\renewcommand{\textbf}[1]{#1}
\renewcommand{\textit}[1]{}
\begin{document}

\title{Moving-Mesh Simulations of Mini-Common Envelope Ejection in Classical Novae}

\author{Nicholas Nelson}
 \email{nelso843@uwm.edu, chang65@uwm.edu}
\author{Philip Chang}%
\affiliation{%
Department of Physics and Astronomy\\
University of Wisconsin - Milwaukee\\
3135 N Maryland Ave\\
Milwaukee, Wisconsin, 53211
}%
\begin{abstract}
Although well studied, our understanding of the mass ejection mechanisms of cataclysmic variables remains incomplete.  Recent work suggests that binary interaction plays an important role in driving and shaping this mass ejection and may affect the long-term evolution of the system.  In this paper, we perform a three-dimensional moving-mesh hydrodynamic simulation of a cataclysmic variable system to study the effect of binary interaction on mass ejection.  We find that once the \textit{nearly hydrostatic} flow crosses the ${\rm L}_1$ Lagrange point, the material \textit{becomes hydrodynamic and} is ejected roughly isotropically.  This can be seen in a roughly spherical distribution of the ejecta at large radii.  We also show that the ${\rm L}_2$ Lagrange point is not important in the ejection of mass, contrary to the assumption in some previous work in this area.  Finally, we find that the specific angular momentum of the ejected material is larger than its initial specific angular momentum.  This enhanced angular momentum ejection likely affects \textit{how} the long-term evolution of the binary system.
\end{abstract}

\maketitle

\section{Introduction}
Cataclysmic Variables are binary systems that consist of a white dwarf (WD) that accretes matter from a secondary star.  As the mass piles up on the WD, the layer of accreted material can become sufficiently hot and dense to trigger unstable nuclear burning.  The resulting rapid expansion and expulsion of the accreted material produces a bright transient event known as a classical nova (for a review see \citep{1978ARA&A..16..171G,2021ARA&A..59..391C} and references therein).

Despite being well studied, the mass ejection mechanisms of these systems are not fully understood.  While the root cause is the thermonuclear runaway, the precise mechanism driving mass loss is not clear.  For instance, the mass ejection may be precipitated by the rapid injection of heat leading to impulsive mass loss \citep{2018ApJ...853...27M, 2012ApJ...761...34L, 1978ApJ...226..186S}, a radiatively driven wind \citep{2016MNRAS.458.1214Q, 1976MNRAS.175..305B, 1966MNRAS.132..317F, 1994ApJ...437..802K}, or more intriguingly, by the action of the secondary in a mini-common-envelope ejection episode \citep{1980MNRAS.191..933M, 1985ApJ...294..263M, 1990ApJ...356..250L}.  \citet{2021ARA&A..59..391C} contains detailed descriptions of these different ejection mechanisms and their potential flaws, which we summarize here.  Firstly, impulsive ejection may be ineffective in ejecting the entire accreted envelope.  Secondly, the radiative driven wind may live or die on the interplay between different energy transport mechanisms (e.g., convection vs radiation) and the availability of super-Eddington energy sources \citep{2016MNRAS.458.1214Q}.  And finally, mini-common-envelope ejection may or may not be important relative to the other two mechanisms.

Recently, \citet{2022ApJ...938...31S} argue using 1-D MESA calculations that while radiation pressure driven winds are successfully launched in WDs undergoing unstable nuclear burning on the surface, the sonic point of the wind is well outside of the Roche radius of the WD.  Hence, they argue that the binary companion can help shape and drive much of the initial (and majority) of the mass ejection. Only at late times, when the sonic point retreats below the Roche radius does radiation pressure drive a fast wind that is not heavily \textbf{influenced} by the binary companion.   

The influence of a binary companion on driving mass loss in a nova may increase the angular momentum loss experienced by the binary which influences its evolution  \citep{2024ApJ...977...34T, 2011ApJS..194...28K, 2016MNRAS.455L..16S, 1991A&A...246...84L}.  For example, \citet{2024ApJ...977...34T} compute the evolution of the binary under the influence of enhanced angular momentum loss.  They find that the rate of angular momentum loss due to binary interaction is significantly higher than that caused by a fast wind, which causes the orbit to shrink.  The companion cannot quickly adjust to this change, causing it to fill its Roche lobe and increase the mass transfer rate from companion to WD.  This expands the orbit again in a cyclical process, eventually leading to the binary becoming detached.  However, they prescribed the specific angular momentum loss as mass ejection through the ${\rm L}_2$ Lagrange point, but it is not clear if this is the correct prescription.
 
An obvious next step is to study the mass ejection in classical novae using 3-D simulations.  In this paper, we carry out the first such study.  Using high-resolution moving-mesh simulations, we study how the ejection process occurs, the shape of the ejecta, and the amount of angular momentum that is carried off by this process.  

This paper is organized as follows.  In Section \ref{sec:numerical}, we describe our numerical code, the initial conditions, and our energy injection scheme to model the effects of runaway nuclear burning.  We then describe our results in \S \ref{sec:results}.  We discuss the implications of these results and close with a summary and discussion of future directions in \S \ref{sec:discussion}.  


\section{Numerical Setup}\label{sec:numerical}
We use \changaMM, a moving-mesh hydrodynamic solver for \changa \cite{Chang+17}, which is based on the arbitrary Lagrangian-Eulerian (ALE) scheme developed by \citet{2010MNRAS.401..791S}.  \changaMM differs from the ALE scheme with a different approach to generating the Voronoi mesh, the use of conserved variables rather than primitives to compute the face states.  We refer readers seeking a more detailed description of \changaMM to \cite{Chang+17,2019MNRAS.486.5809P}.  

We model a WD as a dark matter particle with a mass of 1$M_\odot$ and a softening radius of 0.05$R_\odot$.  While this is substantially larger than a typical WD, it is substantially smaller than the Roche radius. For this system, the Roche radius for the WD is 0.532 $R_\odot$ and the ${\rm L}_1$ Lagrange point is at 0.621 $R_\odot$.   This ``enlarged'' WD is to ensure that the time steps for gas inside the softening radius are not prohibitively small.  We then compute a hydrostatic constant entropy envelope of $10^{-3}\,M_\odot$ and a radius of $0.1 R_\odot$ with an ideal gas equation of state.   A classical nova will ignite the envelope at the surface of the WD, but modeling this initial expansion is prohibitively computationally expensive. Hence, we adopt the assumption that a substantial amount of nuclear burning has taken place and that the envelope has expanded hydrostatically to this radius, which remains inside the Roche radius. In any case, we model this envelope with 100K mesh generating points.  Surrounding this envelope, we include an atmosphere with density of $10^{-10}\,{\rm g\,cm^{-3}}$ with about 250K mesh generating points.  Grid points between the envelope and the atmosphere are distributed as described in \cite{2023MNRAS.526.5365V}.  The entire simulation box is 40 $R_\odot$ in size.

\textbf{
The envelope is not modeled with any spin as we do not believe it to be important in this study.  For a maximally spinning envelope, the specific angular momentum, $\ell$, is given by:
\begin{equation}
    \ell=\sqrt{GM_{\rm WD}R_{\rm WD}}.
\end{equation}
For constant $\ell$, the initial rotation rate of the envelope, $\Omega_i$, is:
\begin{equation}
    \Omega_i=\sqrt{\frac{R_{\rm WD}}{R_i}}\sqrt{\frac{GM_{\rm WD}}{R_i^3}}
\end{equation}
which is the orbital angular velocity at $R_i$ modified by the factor $\sqrt{R_{\rm WD}/R_i}$.  For $R_{\rm WD}=0.05R_\odot$ and an envelope of $R_{\rm env}=0.1R_\odot$ this factor is $\sim0.71$.  When the envelope extends to the Roche radius, $R_{Roche}=0.523R_\odot$, this factor is $\sim0.31$.  To be entirely supported by rotation, the envelope would need $R\Omega^2$.  At $R_{\rm env}$, this would be $\sim1/2$ of what is required.  At $R_{Roche}$, this would be $\sim1/100$ of what is required.  This assumes that the envelope is maximally rotating at the surface of the WD, which is highly unlikely as magnetic fields between the envelope and WD will quickly dampen this differential rotation.  We therefore claim that initial rotation of the envelope is not important to model.
}

To model the continued injection of energy into the envelope via nuclear burning, we use a constant specific heating rate of $\epsilon = 10^{12}\,{\rm ergs\,g^{-1}\,s^{-1}}$. This specific heating rate gives a total heating luminosity of $2\times 10^{42}\,{\rm ergs\,s^{-1}}$ and is similar to the nova outburst energy injection rate.  It ensures that we expand \textit{hydrostatically} \textbf{ slowly (relative to the local dynamical time)} to the ${\rm L}_1$ Lagrange point as suggested by \citep{2022ApJ...938...31S}.  We have experimented with a number of different specific heating rates varying from $10^8$ - $10^{13}\,{\rm ergs\,g^{-1}\,s^{-1}}$.  Significantly larger heating rates will drive a strong wind, whereas smaller heating rates will take a prohibitively long time to evolve.   We also model a companion star of 0.3$M_\odot$ as a dark matter particle and place it in a circular orbit with $R = R_\odot$.  In reality, this companion star would fill its Roche lobe and would have an accretion stream that feeds an accretion disk around the WD, but for the purposes of this study, we ignore these effects.  In practice, ignoring the accretion stream and disk is likely an acceptable approximation as its mass is small compared to the ejecting envelope.  However, this is likely not the case for the Roche filling secondary star. 

\textbf{
We assume a solar abundance for the envelope.  This is because the amount of energy needed to unbind the envelope is small compared to the potential nuclear energy released.  We do not account for mixing of WD material into the envelope. For instance, \citep{2016PASP..128e1001S} showed that up to about 20\% of the composition can be changed due to mixing.  Thus, the assumed solar composition is a fair approximation.
}

\textbf{
The total heating luminosity is consistent with that from Figure 1 of \cite{2022ApJ...938...31S}.  As the effective temperature of the WD goes from $6\times10^5 \,{\rm K} \to 10^4 \,{\rm K}$ with roughly constant luminosity we approximate the change in radius.  Since $L=4\pi R^2\sigma T^4$ is constant, we can write:
\begin{equation}
    \left(\frac{T_i}{T_f}\right)^4=\left(\frac{R_f}{R_i}\right)^2
\end{equation}
and with $T_i/T_f=60$, $R_f/R_i=3600$.  
For $L=10^4L_\odot$ and $T_i=6\times10^5 \,{\rm K}$, we can determine that $R_i=7\times10^8\, {\rm cm}$ or $R_i=0.01R_\odot$, which is consistent with the radius of WDs.  We can then determine the energy outburst, $\dot E$ with the following:
\begin{equation}
    \dot E = \frac{GM_{\rm WD}M_{\rm env}}{R_i}\frac{1}{t_{ob}}.
\end{equation}
For the outburst time, $t_{ob}=4.8\, {\rm days}=4.32\times10^5 \,{\rm s}$, this equation would yield $\dot E=8.82\times10^{41} \,{\rm ergs\, s^{-1}}$ which is consistent with that obtained from our constant specific heating rate.
}

\textbf{
We can ignore the accretion disk as its mass is negligible in comparison to the mass of the envelope.  The mass of the envelope can be given by the following:
\begin{equation}
    M_{\rm env} = \dot M t_{\rm rec}
\end{equation}
where $t_{\rm rec}$ is the classical nova recurrence time.  Similarly, the mass of the disk can be given by:
\begin{equation}
    M_{\rm disk}=\dot M t_{\rm visc}
\end{equation}
where $t_{\rm visc}$ is the viscous timescale of the accretion disk and is given by
\begin{equation}
    t_{\rm visc}=\left(\frac{R}{H}\right)^2\frac{1}{\alpha}t_{\rm dyn}.
\end{equation}
Thus, the ratio of the two masses are
\begin{equation}
    \frac{M_{\rm disk}}{M_{\rm env}}=\frac{t_{\rm visc}}{t_{\rm rec}}.
\end{equation}
For $R/H \sim 10$ and $1/\alpha \sim 10$, we find for a $t_{\rm rec} \sim 10$ yrs \cite{2021ARA&A..59..391C} $M_{\rm disk}\sim 10^{-3}M_{\rm env}$ and is thus negligible.
}


We simulated the above system for \textit{16.7} \textbf{27.78} hours or $10^5$ seconds, which took a wall clock time of 30 days on a single 64 core node.




\section{Results}\label{sec:results}
\begin{figure*}
\includegraphics[width=0.3\textwidth]{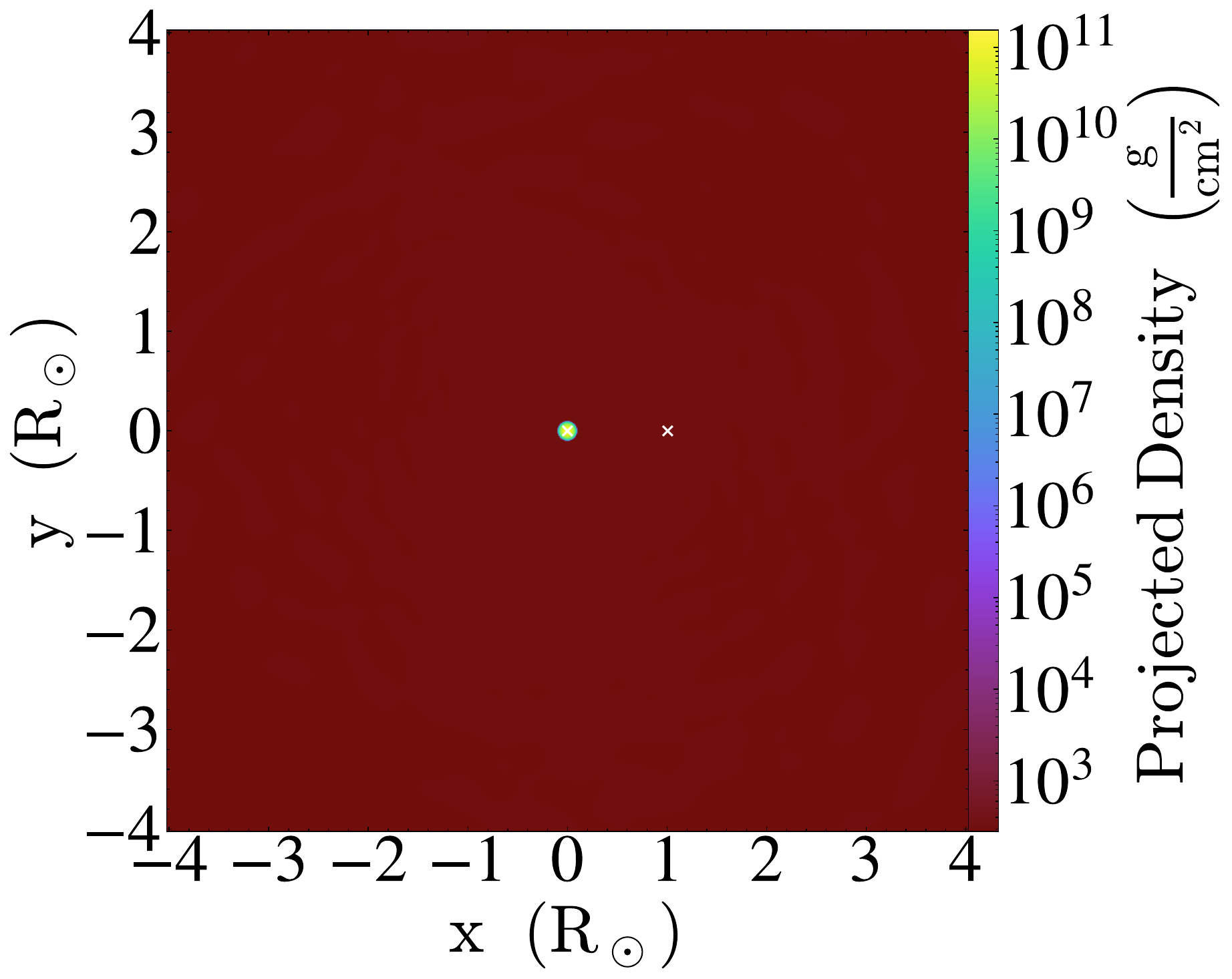}
\includegraphics[width=0.3\textwidth]{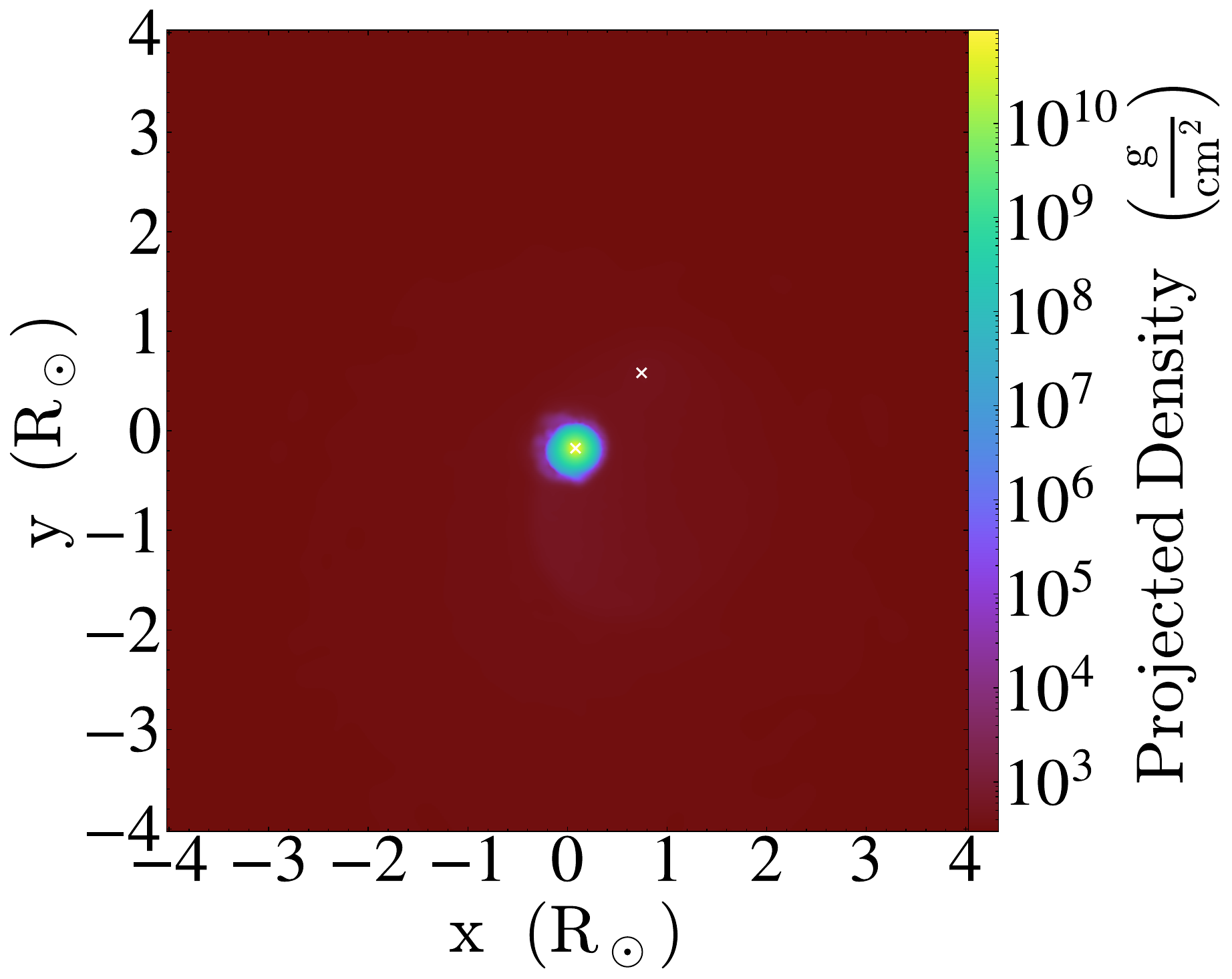}
\includegraphics[width=0.3\textwidth]{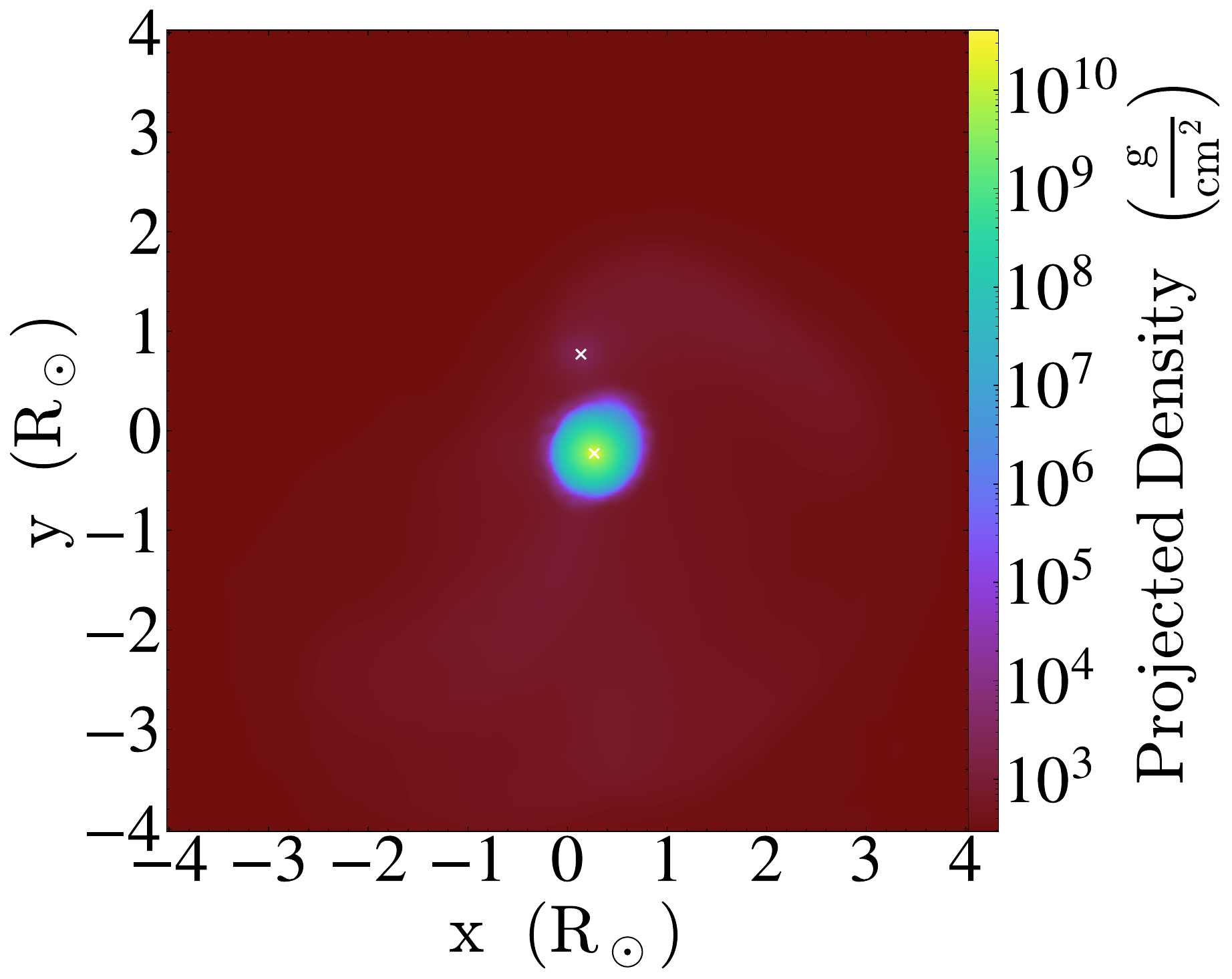}\\
\includegraphics[width=0.3\textwidth]{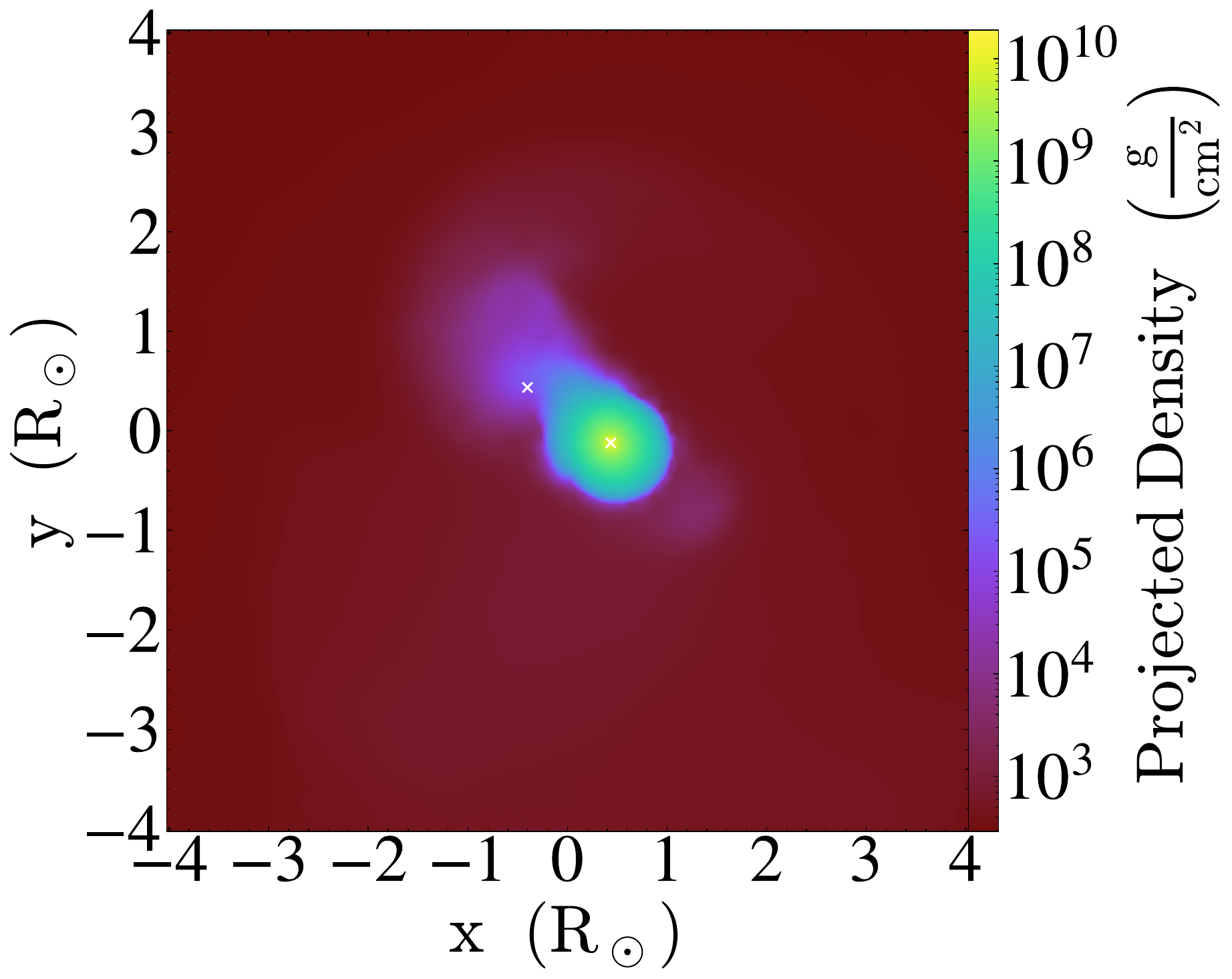}
\includegraphics[width=0.3\textwidth]{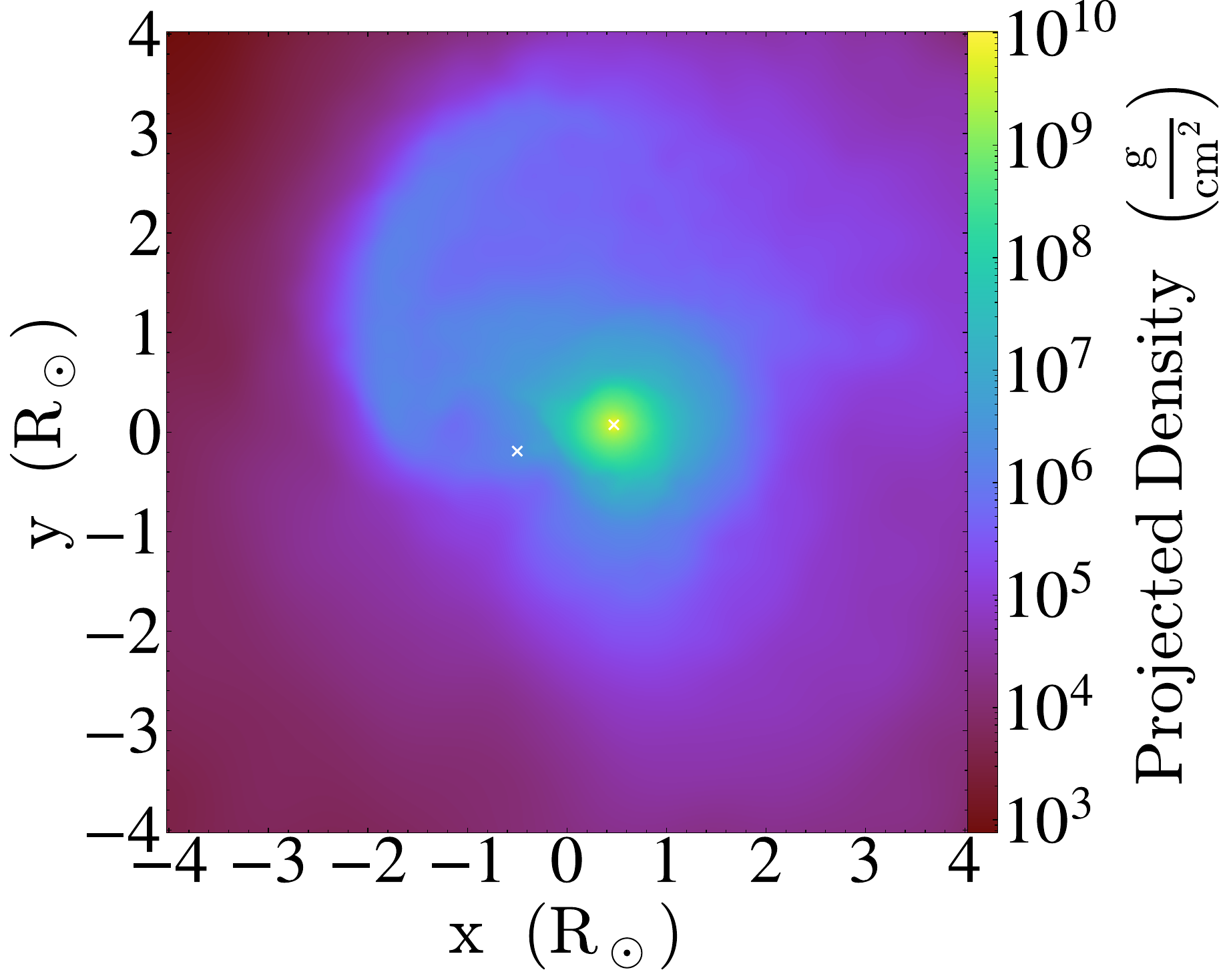}
\includegraphics[width=0.3\textwidth]{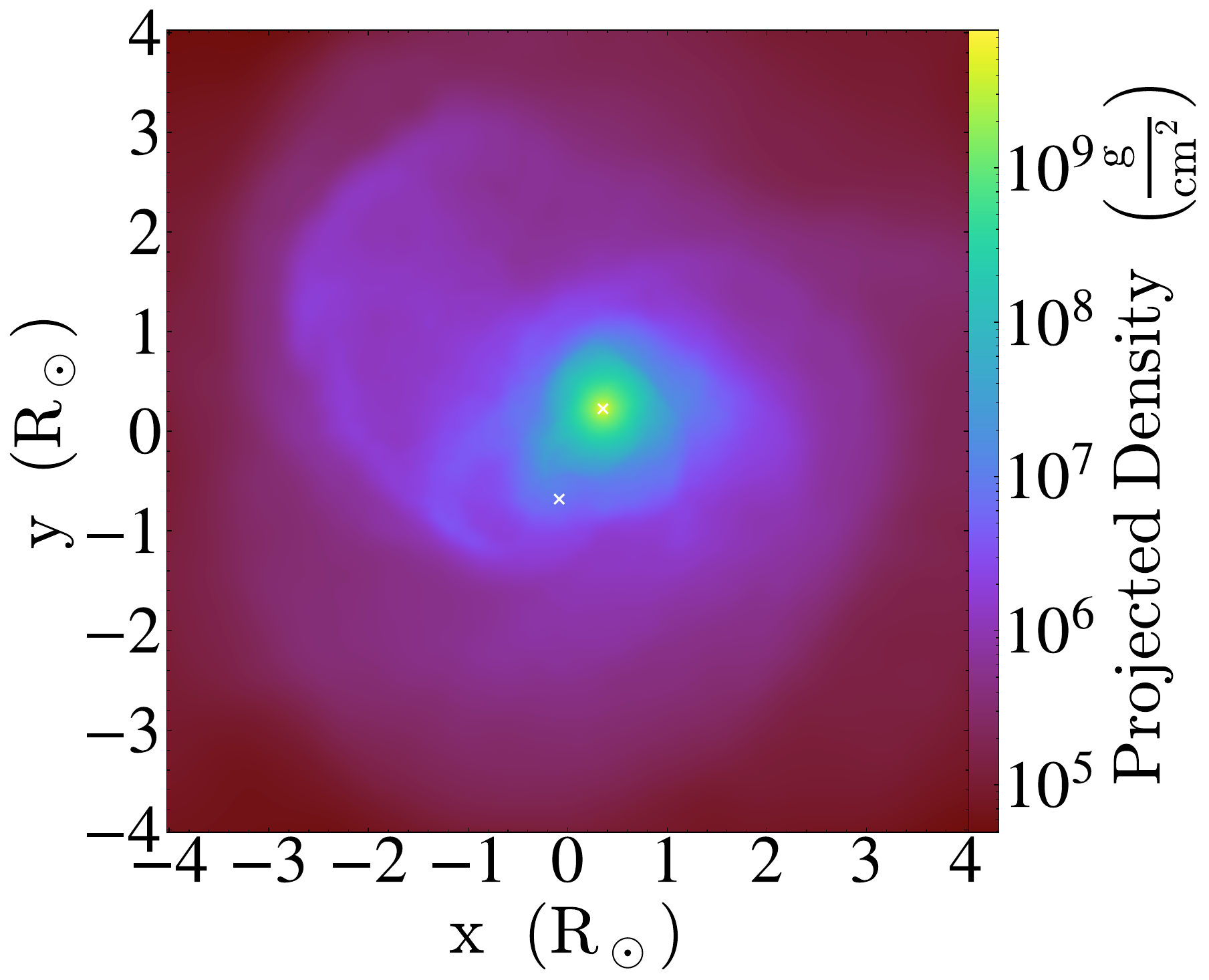}
\caption{\label{fig:projxy} Projected density along the z or orbital axis, at $t = 0$, $2.78$, $5.56$, $8.33$, $11.11$, and $13.89$ hours.}
\end{figure*}

\begin{figure*}
\includegraphics[width=0.3\textwidth]{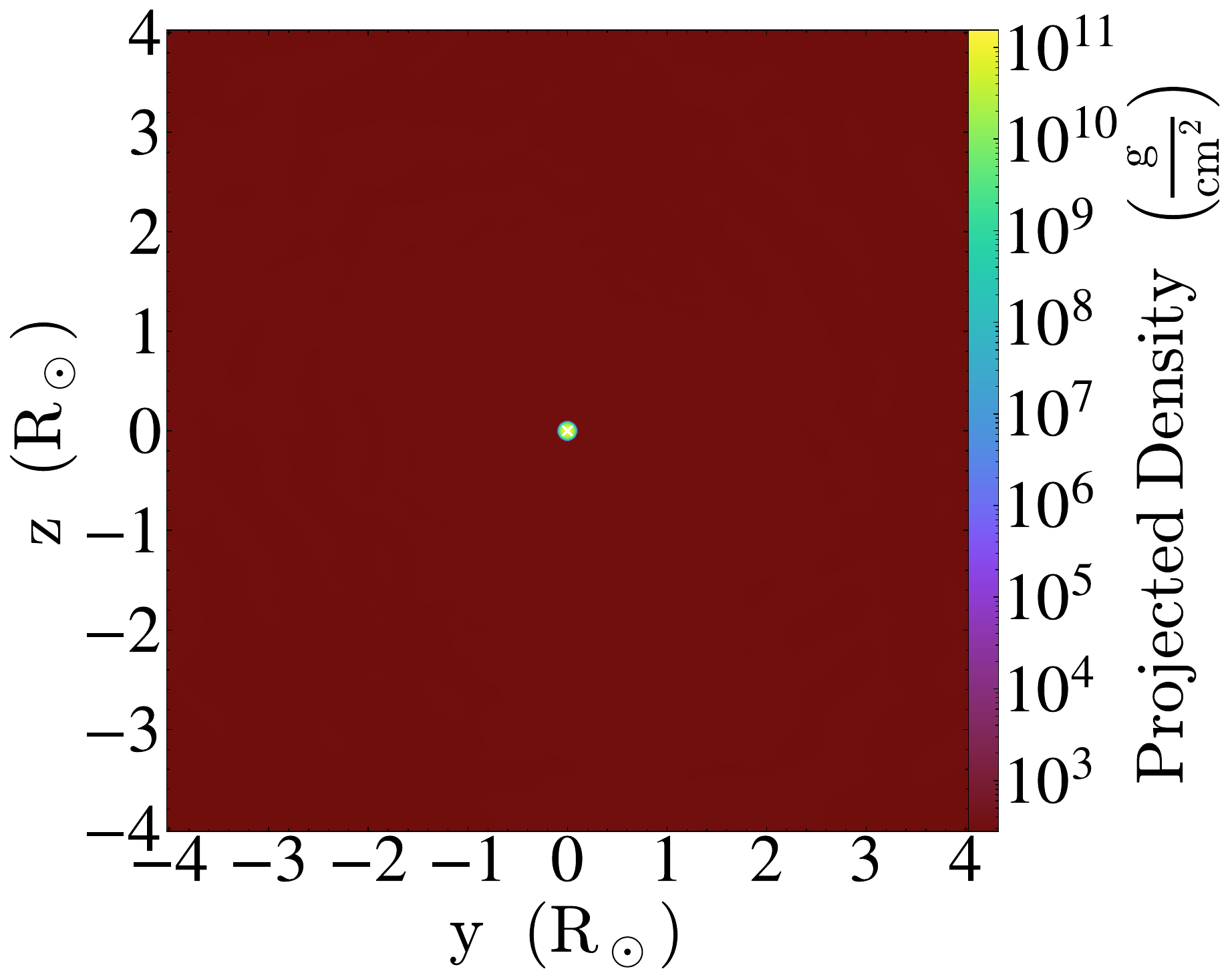}
\includegraphics[width=0.3\textwidth]{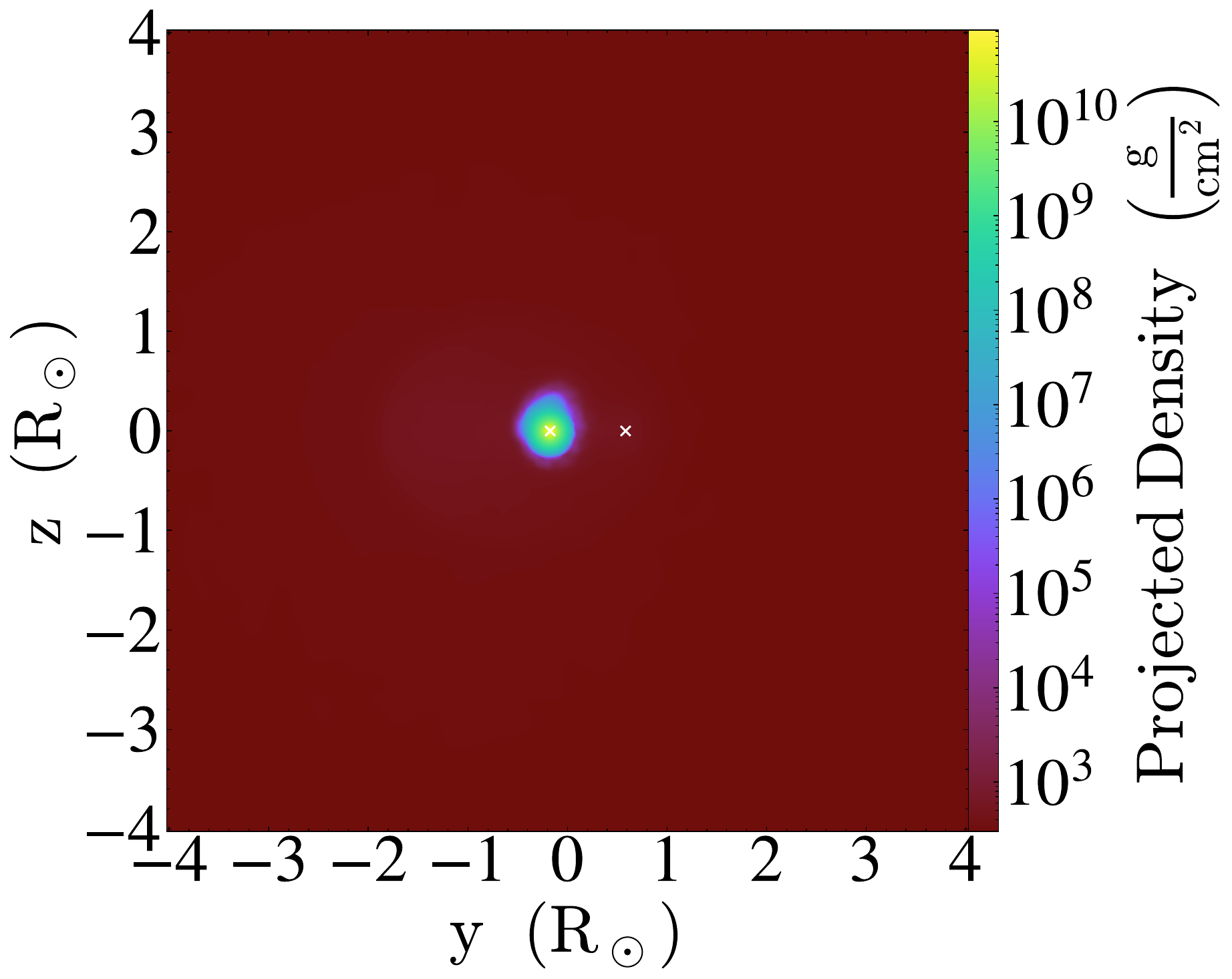}
\includegraphics[width=0.3\textwidth]{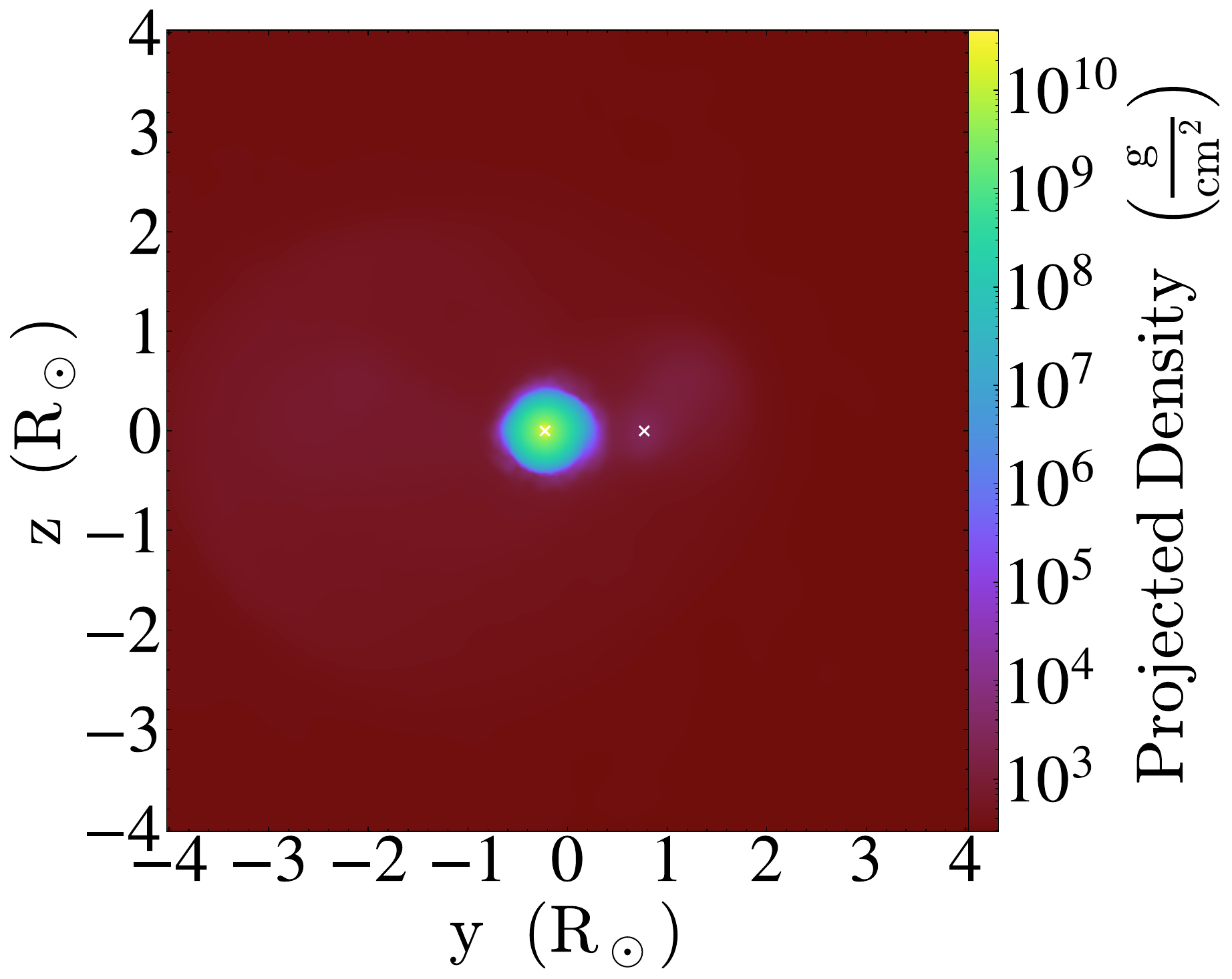}\\
\includegraphics[width=0.3\textwidth]{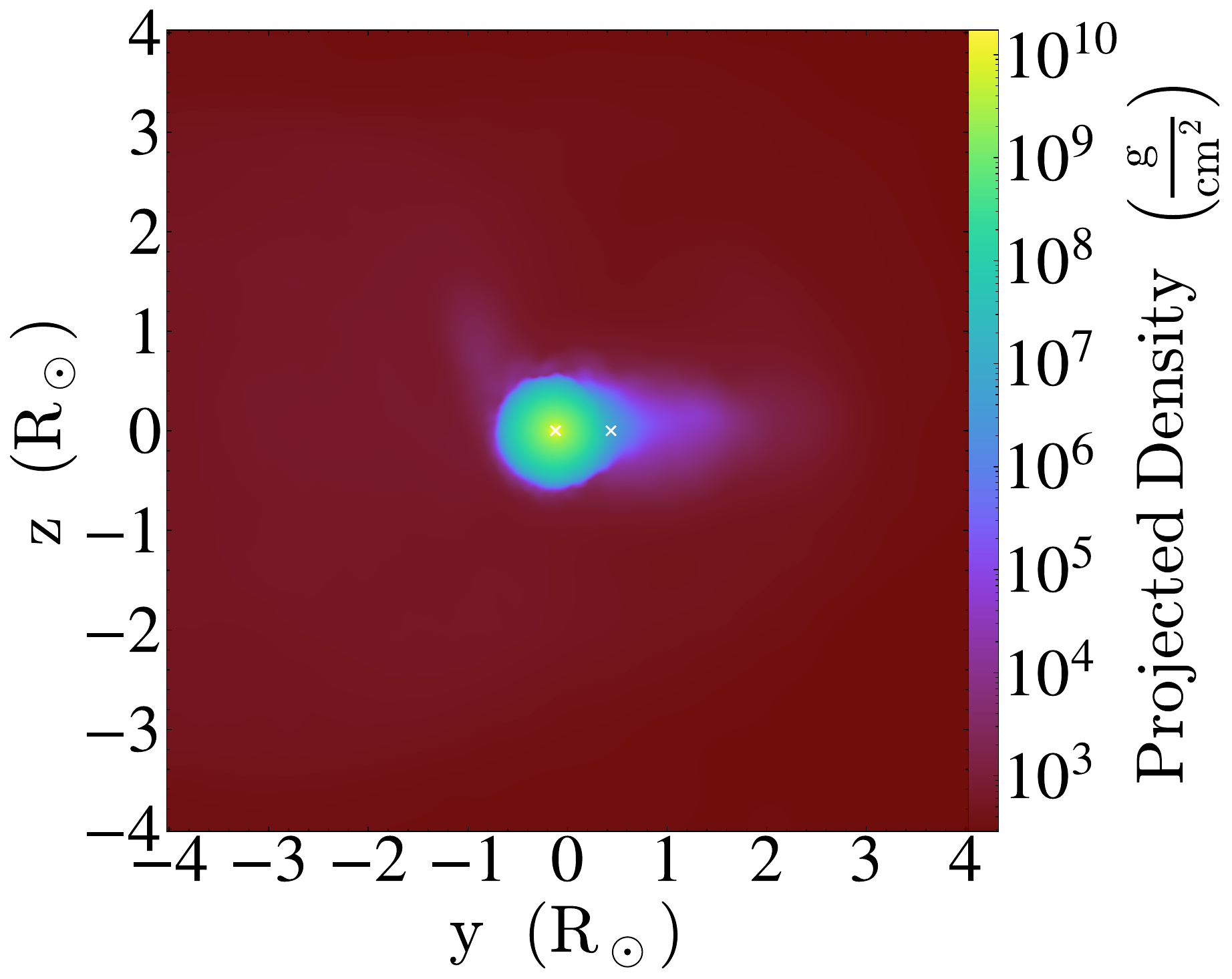}
\includegraphics[width=0.3\textwidth]{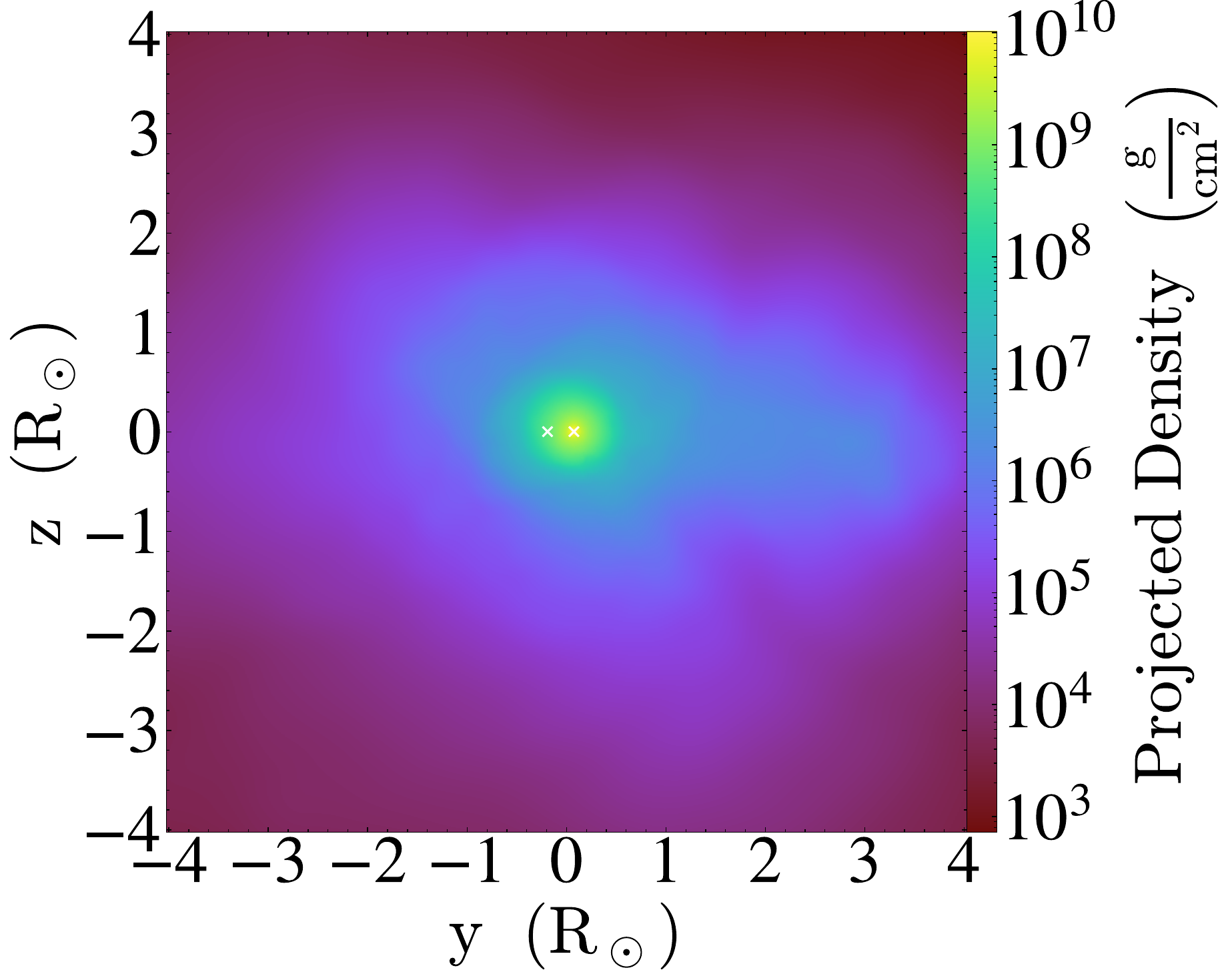}
\includegraphics[width=0.3\textwidth]{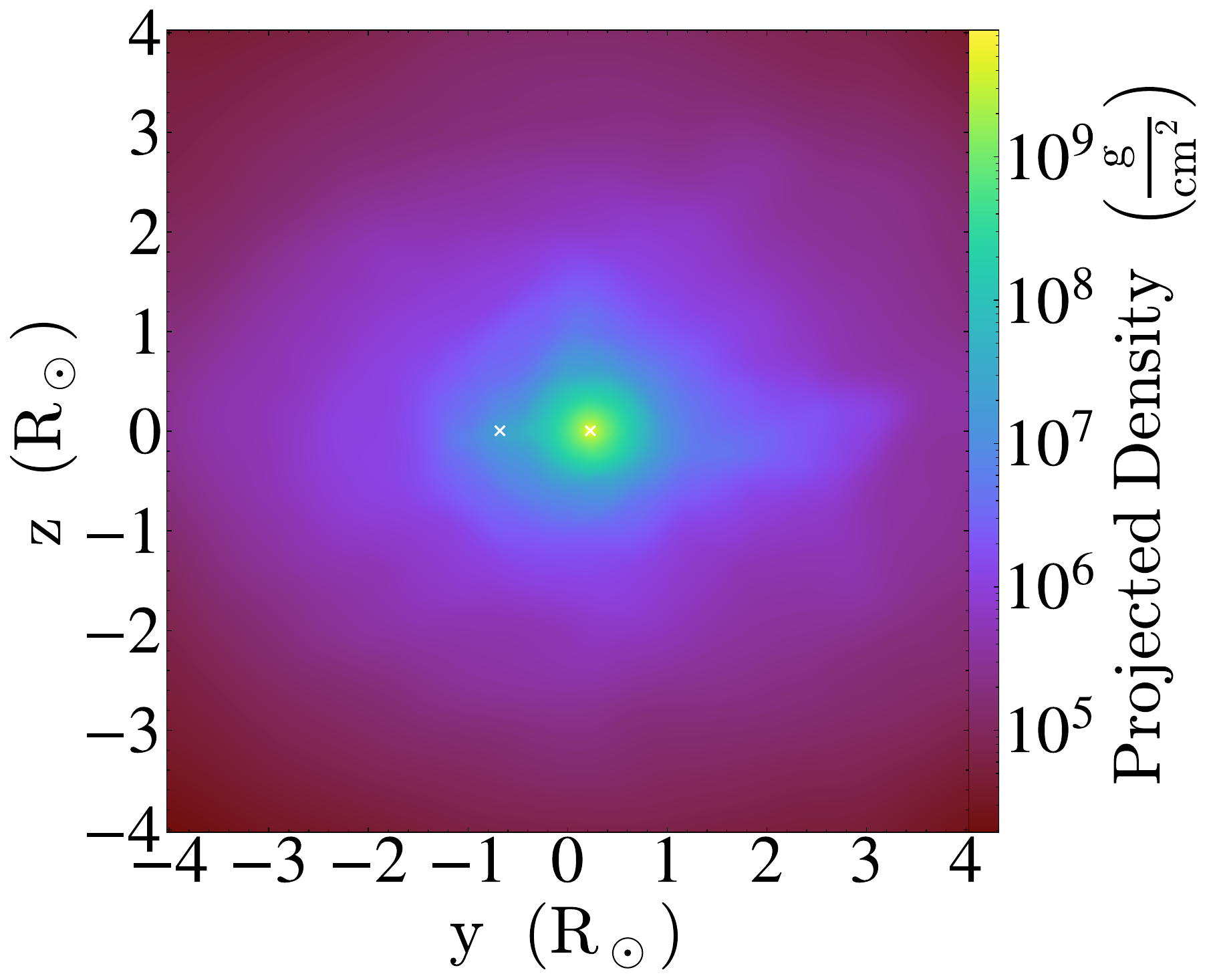}
\caption{\label{fig:projyz} Projected density along the x-axis, in the orbital plane, at $t = 0$, $2.78$, $5.56$, $8.33$, $11.11$, and $13.89$ hours.}
\end{figure*}

\begin{figure*}
\includegraphics[width=0.45\textwidth]{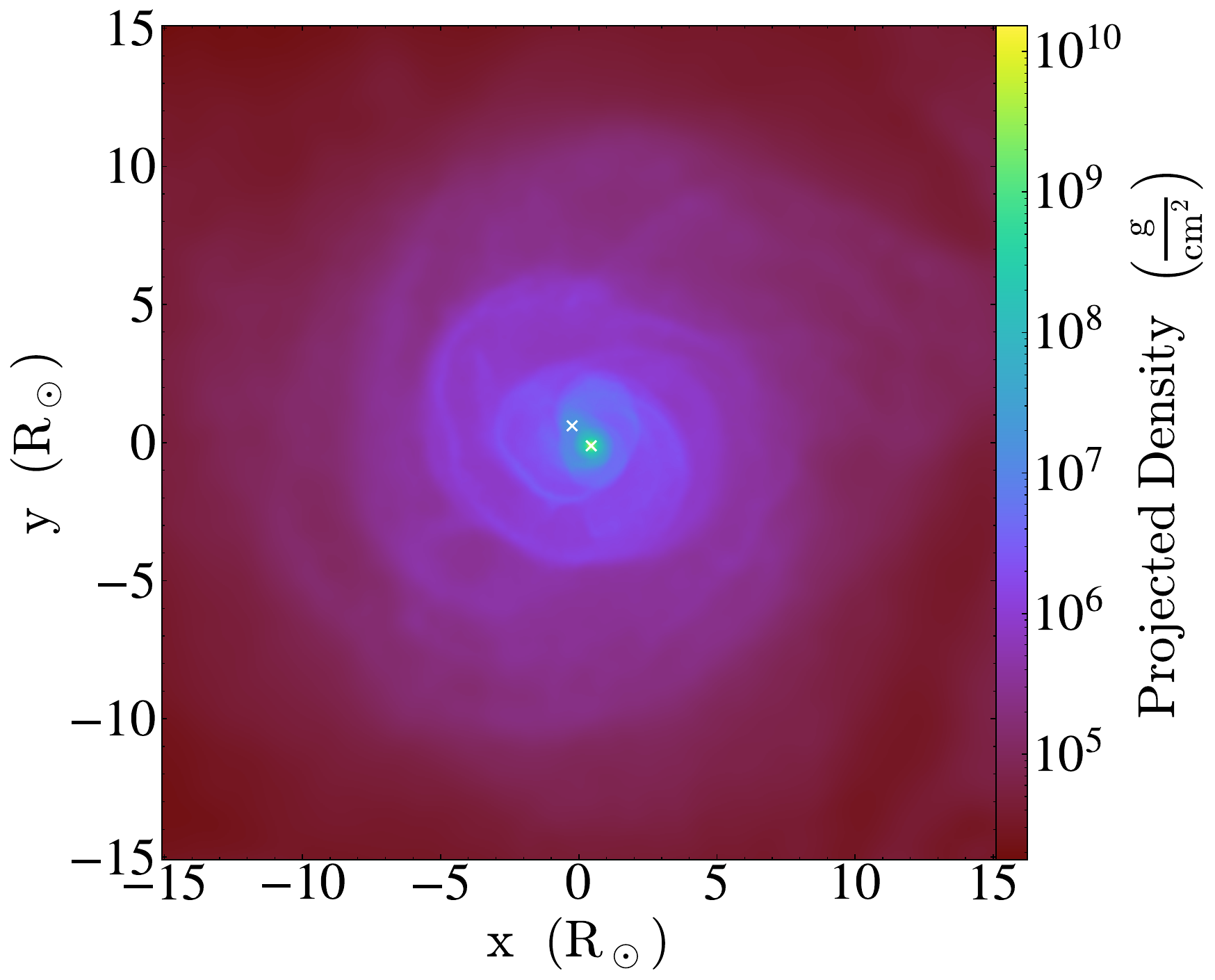}
\includegraphics[width=0.45\textwidth]{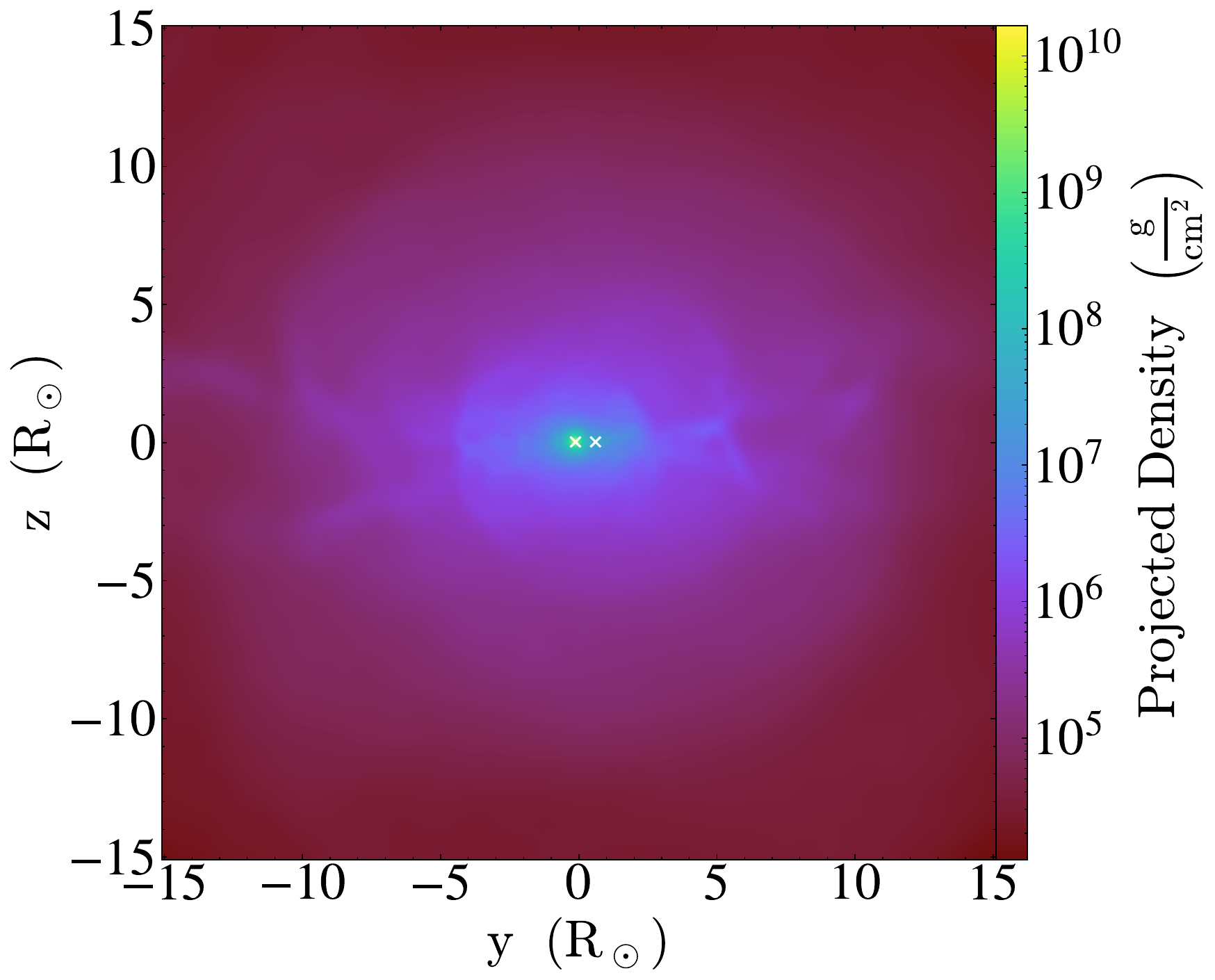}
\caption{\label{fig:bigframe} Projected density along the z- and x-axes at $t = 27.78$ hours.}
\end{figure*}

Figure \ref{fig:projxy} shows the projected density along the z or orbital axis, and Figure \ref{fig:projyz} shows the projected density along the x-axis, which is in the orbital plane, at $t = 0$, $2.78$, $5.56$, $8.33$, $11.11$, and $13.89$ hours.  At $t = 0$, the initial setup of the system is shown.  At $2.78$ hours, the hydrostatic expansion of the envelope of the WD can be seen due to the heating described above.  At $5.56$ hours, the WD has nearly filled its Roche lobe and small amounts of the envelope are beginning to flow across the ${\rm L}_1$ point towards the companion.  This is more apparent at $8.33$ hours, where the WD has fully filled its Roche lobe and the envelope is flowing through the ${\rm L}_1$ Lagrangian point.  This leads to ejection of material through interaction with the secondary, as shown in the last two frames at $11.11$ hours and at $13.89$ hours.  In the last two frames, the ejected material around the system appears to be in a roughly spherical distribution.  \textbf{Figure \ref{fig:bigframe} shows the projected density along the z- and x-axes at $t=27.78$ hours, which is the final frame of our simulation.  This figure qualitatively agrees with the final two plots of Figures \ref{fig:projxy} and \ref{fig:projyz}, emphasizing the ejected material's roughly spherical distribution.}

\subsection{Spherically averaged quantities}

\begin{figure}
\includegraphics[width=0.45\textwidth]{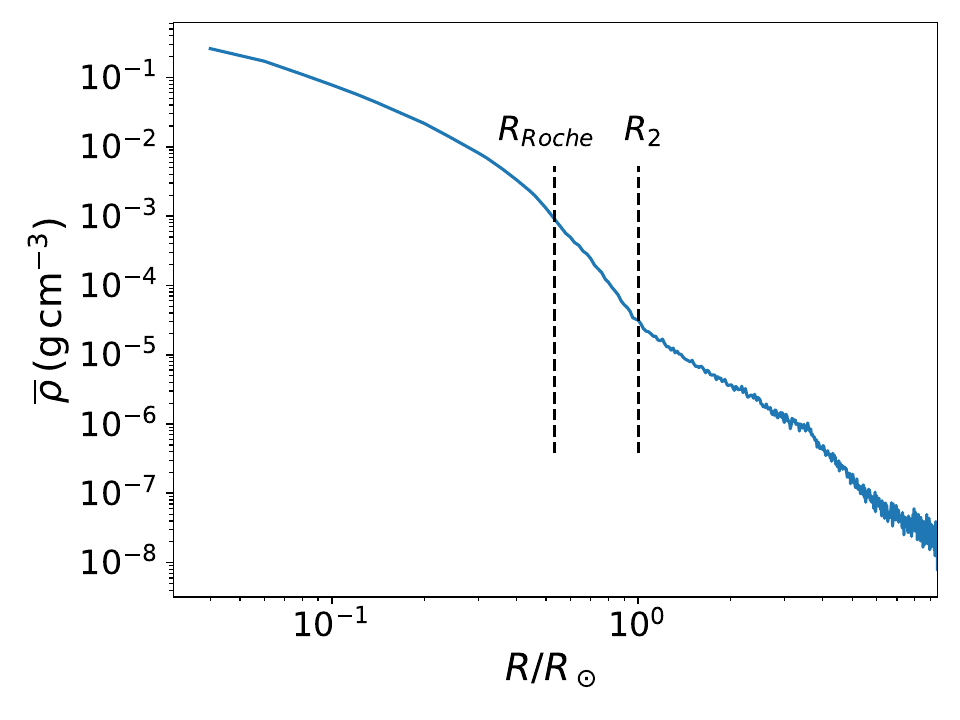}
\caption{\label{fig:density_variance} Spherically averaged gas density as a function of radius with respect to the WD at $t=14.17$ hours. The density behaves roughly as a broken power law, with breaks at $R=R_{Roche}$ and at the companion ($R=R_2$).}
\end{figure}

Motivated by the rough spherical distribution of the ejected material in Figures \ref{fig:projxy}, \ref{fig:projyz}, \textbf{and \ref{fig:bigframe}}, we plot the spherically averaged density at t=14.17 hours in Figure \ref{fig:density_variance}. \textit{, which is near the end point of our simulation.}



\begin{figure}
    \includegraphics[width=0.45\textwidth]{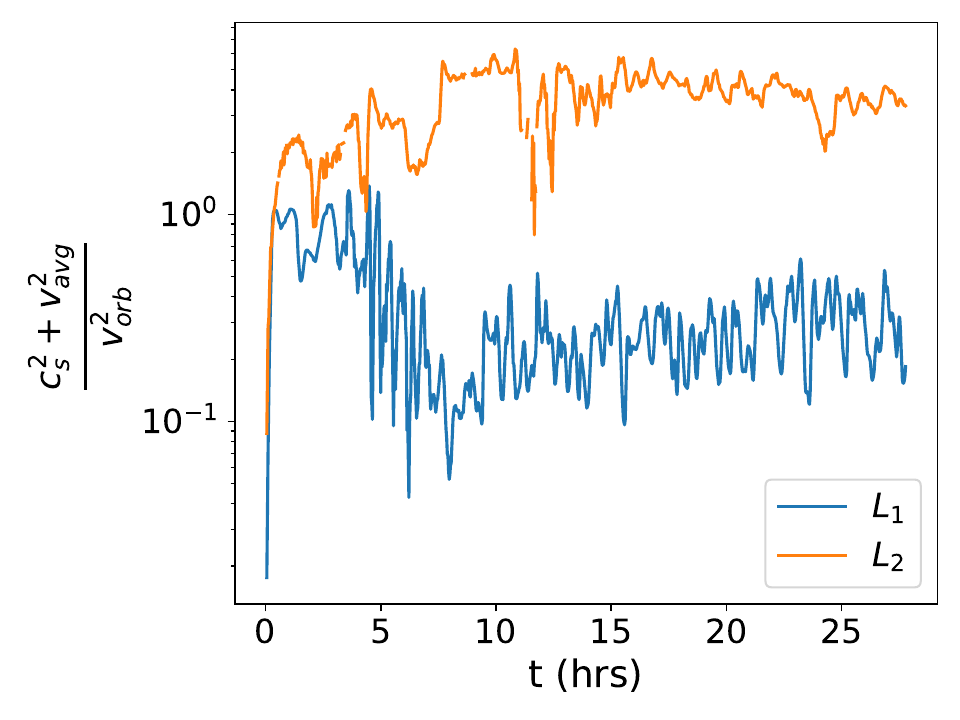}\\
    \includegraphics[width=0.45\textwidth]{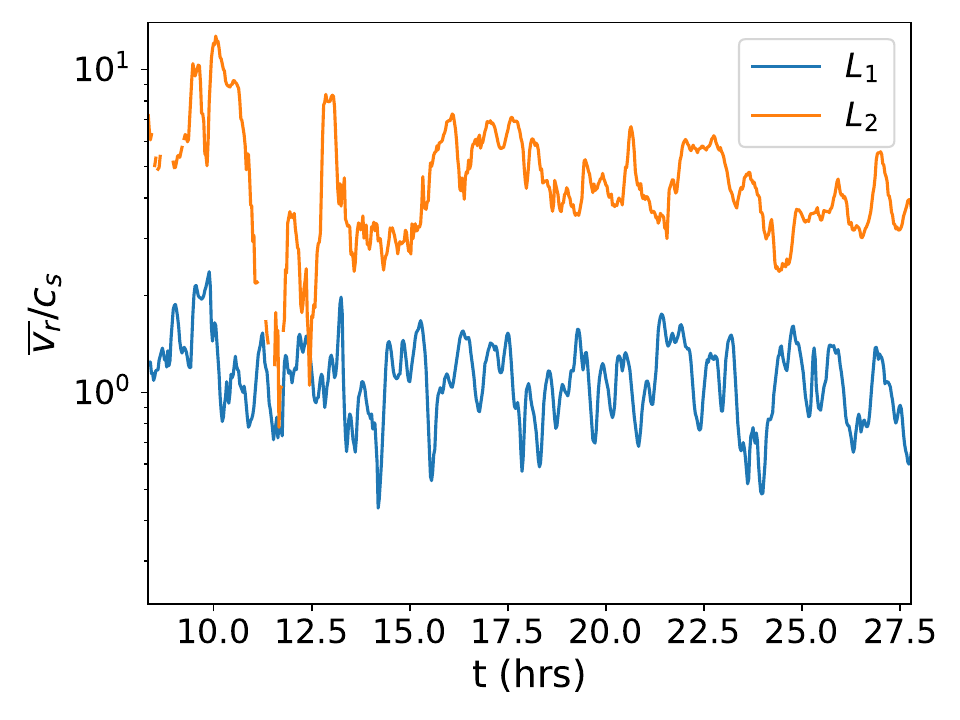}
    \caption{\label{fig:energy_mach} Ratio of kinetic and internal energies to the potential energy at the ${\rm L}_1$ and ${\rm L}_2$ Lagrange points (top plot).  Mach number of the flow at the ${\rm L}_1$ and ${\rm L}_2$ Lagrange points (bottom plot). The above plots suggest that the flow around ${\rm L}_1$ is \textit{hydrostatic} \textbf{ driven by thermal and kinetic energy} while the flow around ${\rm L}_2$ is \textit{hydrodynamic} \textbf{ dominated by kinetic energy}.}
\end{figure}
    
Figures \ref{fig:projxy} and \ref{fig:projyz} demonstrate that the secondary is a crucial ingredient in the ejection of the expanding envelope in line with \cite{2022ApJ...938...31S}.  Motivated by this, we now look at how the flow changes around the various Lagrange points.  In the top plot of Figure \ref{fig:energy_mach}, we show the ratio of kinetic and internal energy to the gravitational potential energy, $(c_s^2 + v^2)/v_{\rm orb}^2$  around ${\rm L}_1$ and ${\rm L}_2$ Lagrange points. From Figure \ref{fig:projxy}, we note that a substantial flow begins to pour across the ${\rm L}_1$ Lagrange point around 5-8 hours.  Here we note that after $t\approx 6$ hours, the ratio of kinetic and internal energy to potential energy is $<1$.  \textit{The ratio is around 0.2-0.4 after this time, indicating that the material is bound and expands hydrostatically.}\textbf{ The ratio is around 0.2-0.4 after this time, indicating that the material is bound to the system.} In contrast, the flow around ${\rm L}_2$ is always greater than unity and in many cases substantially so.  \textit{Thus, the fluid around ${\rm L}_2$ is not bound and is not hydrostatic.}\textbf{  Thus, the fluid around ${\rm L}_2$ is not bound to and can escape from the system.}  Instead, it suggests that the fluid around ${\rm L}_2$ flows outward.

The bottom plot of Figure \ref{fig:energy_mach} shows the mach number of the flow around the ${\rm L}_1$ and ${\rm L}_2$ Lagrange points.  Here we note that the mach number is around unity for ${\rm L}_1$, which suggests that the material is supported by a combination of kinetic and thermal energies.  
Contrast this with the flow around ${\rm L}_2$, where the mach number is substantially greater than unity for the most part, which again suggests that pressure plays a minor role and the material is flowing outward. Combined with the result that the kinetic and internal energies are larger than the potential energy for the ${\rm L}_2$ Lagrange point, as shown in Figure \ref{fig:energy_mach}, we can conclude that the material should move ballistically.  


\begin{figure}
    \includegraphics[width=0.45\textwidth]{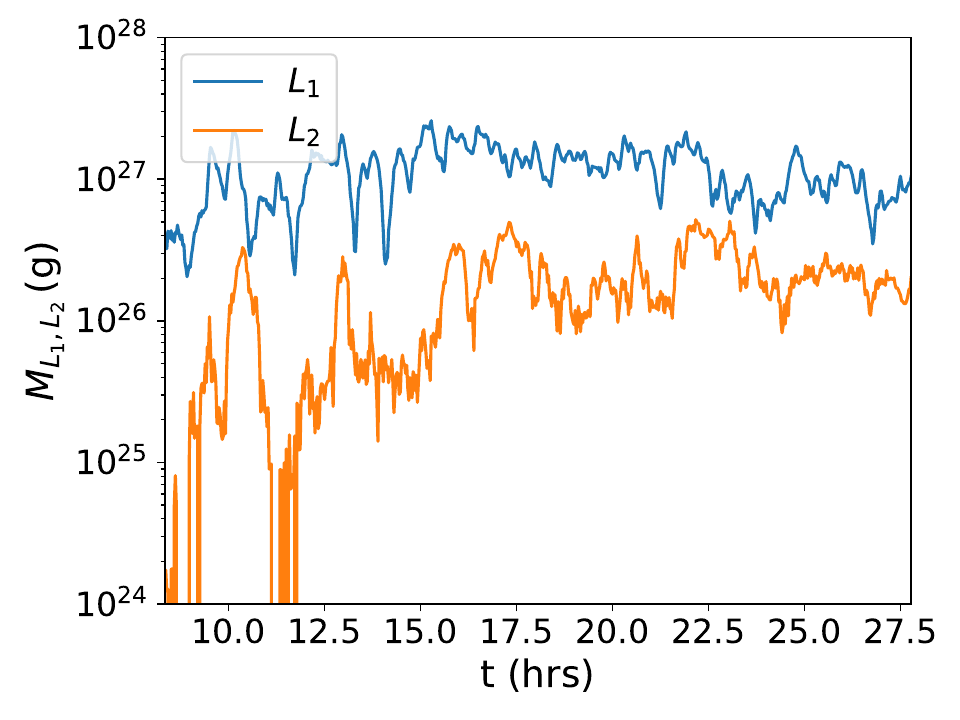}
    \caption{\label{fig:mass_L1,L2} Mass \textbf{ near} the ${\rm L}_1$ and ${\rm L}_2$ points as a function of time. Note that the mass around ${\rm L}_1$ is substantially larger than the mass around ${\rm L}_2$.  The ballistic nature of the flow at this point coupled with low mass implies that the ${\rm L}_2$ point is not significant.}
\end{figure}

\begin{figure}
    \includegraphics[width=0.45\textwidth]{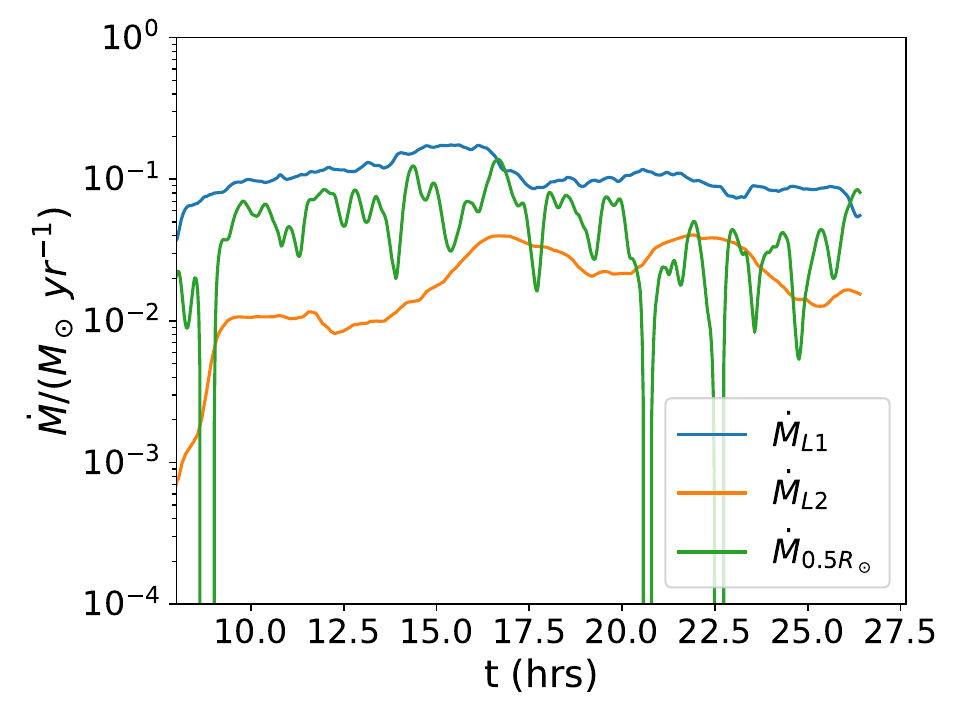}
    \caption{\label{fig:massflux} \textbf{ Mass flux in the shells around the ${\rm L}_1$ Lagrange point, the ${\rm L}_2$ point, and $R=0.5R_\odot$, each measured with respect to the companion star with a thickness of $0.05R_\odot$.  Note that the drop-offs for $\dot M_{0.5R_\odot}$ are due to noise caused by the movement of the companion star.} }
\end{figure}
    
As the flow appears ballistic at the ${\rm L}_2$ Lagrange point, it suggests that this point, which is determined by a combination of gravity and orbital motion, may not be a meaningful point in the flow.  Contrast this with the ${\rm L}_1$ Lagrange point where gravity \textbf{, pressure, and kinetic energy all} appear\textit{s} to play \textit{an} important role\textbf{s} \textit{due to its locally hydrostatic nature}.  We can see further evidence of the minimal effects of the ${\rm L}_2$ Lagrange point in Figure \ref{fig:mass_L1,L2}, where we calculate the mass of gas in a $0.1R_\odot$ sphere around the ${\rm L}_1$ and ${\rm L}_2$ Lagrange points. In nearly the entire run, the mass in gas around the ${\rm L}_2$ Lagrange point is much smaller than the mass in gas around the ${\rm L}_1$ Lagrange point.  As this material obeys continuity coupled with the fact that the flow appears ballistic around the ${\rm L}_2$ Lagrange point, we can conclude that the gas escapes in a region that is much larger than the ${\rm L}_2$ Lagrange point, e.g., $4\pi$ steradians around the secondary. \textbf{To reinforce this point, Figure \ref{fig:massflux} calculates the mass flux through different sections of the $R=0.5{\rm R}_\odot$ spherical shell with thickness $0.05 {\rm R}_\odot$ centered around the secondary. The region of $\dot M_{{\rm L}1}$ is a $30^{\circ}$ spherical section centered toward the ${\rm L}_1$ point.  Similarly, the region of $\dot M_{{\rm L}2}$ is a $30^{\circ}$ spherical section centered toward the ${\rm L}_2$ point.  The final region is the remainder of this spherical shell, which we define as $\dot M_{0.5{\rm R}_\odot}$ and refer to as the remaining region.  Figure \ref{fig:massflux} shows that the mass flux that through the ${\rm L}_1$ region will flow out through both the ${\rm L}_2$ and remaining regions.  This figure also shows that, largely, most of the outflowing mass moves through the remaining region.  In all cases, the mass inflow through the ${\rm L}_1$ region is much larger than the mass outflow through the ${\rm L}_2$ region.}

\begin{figure}
    \includegraphics[width=0.45\textwidth]{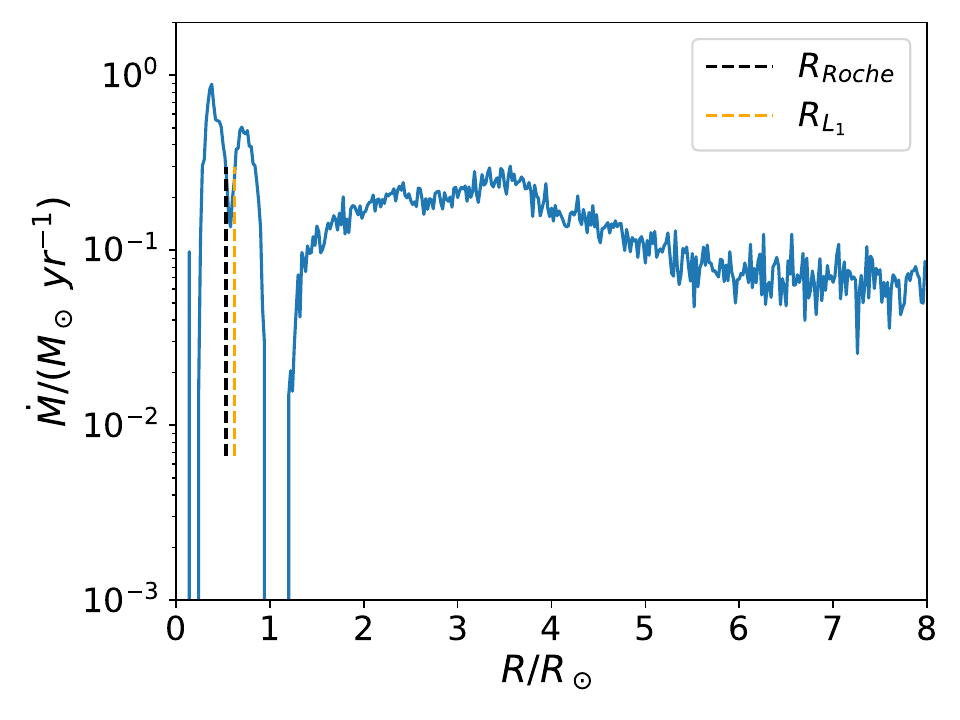}\\
    \includegraphics[width=0.45\textwidth]{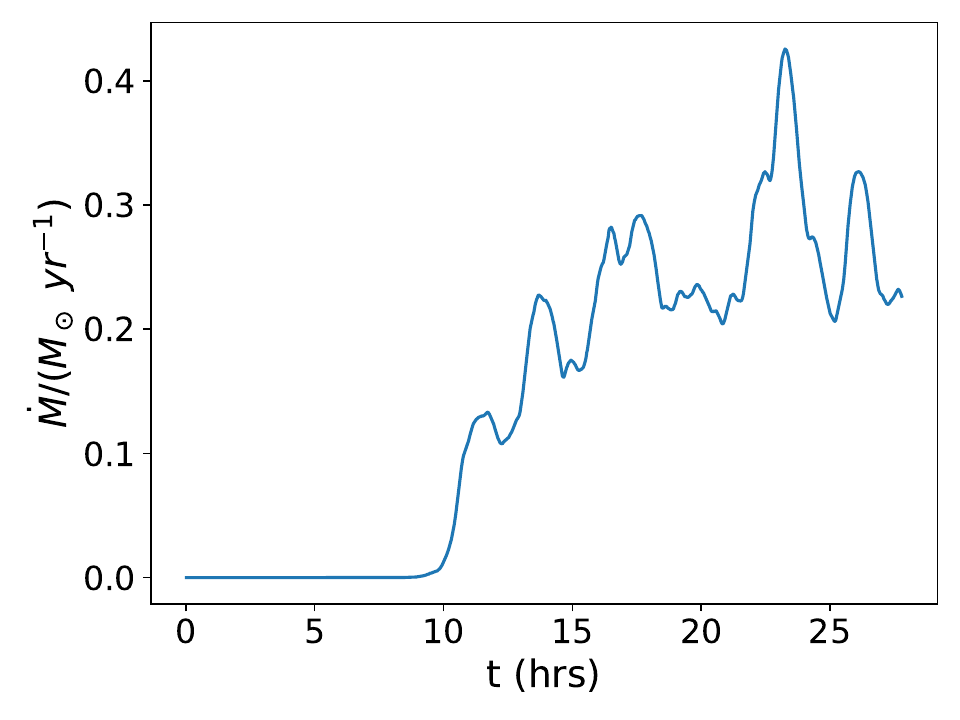}

    \caption{\label{fig:mdot} $\dot M$ as a function of radius at $t=14.17$ hours (top plot). $\dot M$ as a function of time at $r=3R_\odot$ (bottom plot).  Well outside the binary, $\dot M \sim 0.1\,M_\odot {\rm yr^{-1}}$.  From the bottom plot, we can approximate a mass loss rate of $\dot M \approx 0.2-0.3\,M_\odot {\rm yr^{-1}}$.}
    \end{figure}
            
\begin{figure}
    \includegraphics[width=0.45\textwidth]{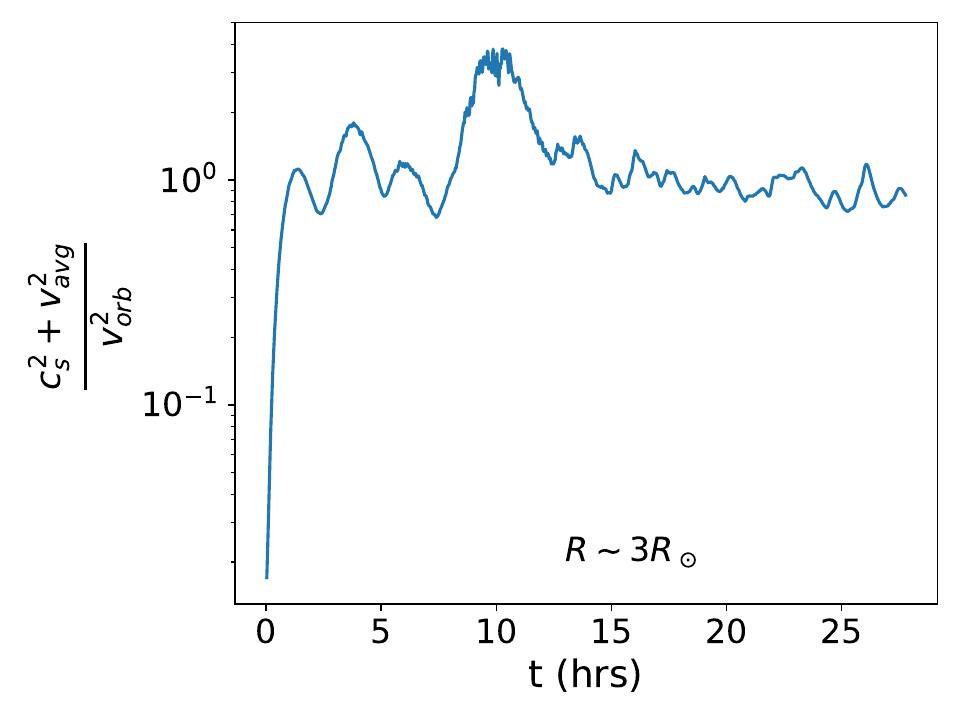}

    \caption{\label{fig:r3energy} \textbf{ Ratio of kinetic and internal energies to the potential energy at the $R=3R_\odot$ spherical shell.}}
    \end{figure}

In the top plot of Figure \ref{fig:mdot}, we plot the spherical mass loss rate, $\dot{M}$, as a function of \textbf{radius} at $t=14.17$ hours. Well outside of the binary, $\dot{M}$ settles around 0.1 $M_{\odot}\,{\rm yr}^{-1}$.  Motivated by this, we then plot $\dot{M}$ at $r = 3 R_\odot$ as a function of \textbf{time}.  In the beginning, there is no mass outflow as the envelope expands.  At 6 hours, the material starts to flow through the ${\rm L}_1$ point as discussed above, and a few hours later this material is ejected from the system and flows through the $r=3R_\odot$ spherical shell.  \textbf{We demonstrate this in Figure \ref{fig:r3energy}, where we plot the ratio of internal and kinetic energies compared to potential energy. As we can see in the plot, the ratio is generally above unity (sometimes substantially so) and, thus, this material is unbound and is expected to leave the system.} This results in a mass loss rate of about $\approx 0.2-0.3 \,M_{\odot}{\rm yr}^{-1}$.  Given the timescales of classical novae of a few days and a mass of the envelope of $10^{-3}\,{M_{\odot}}$, this is reasonable and suggests that the heating rate that we imposed on the problem can approximate novae.

\begin{figure}
    \includegraphics[width=0.45\textwidth]{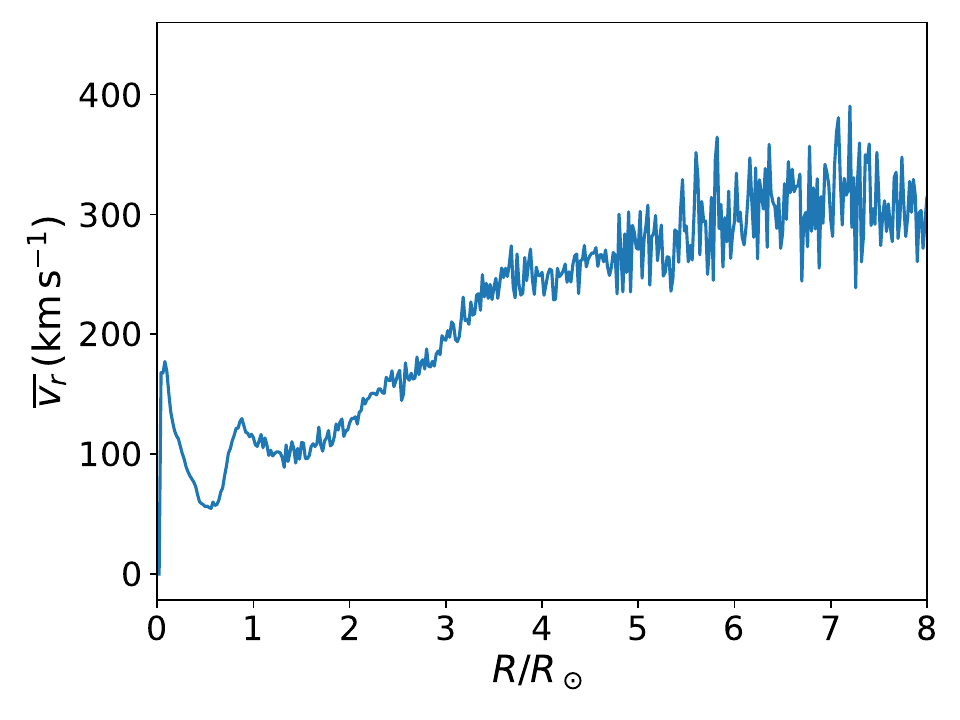}
    \includegraphics[width=0.45\textwidth]{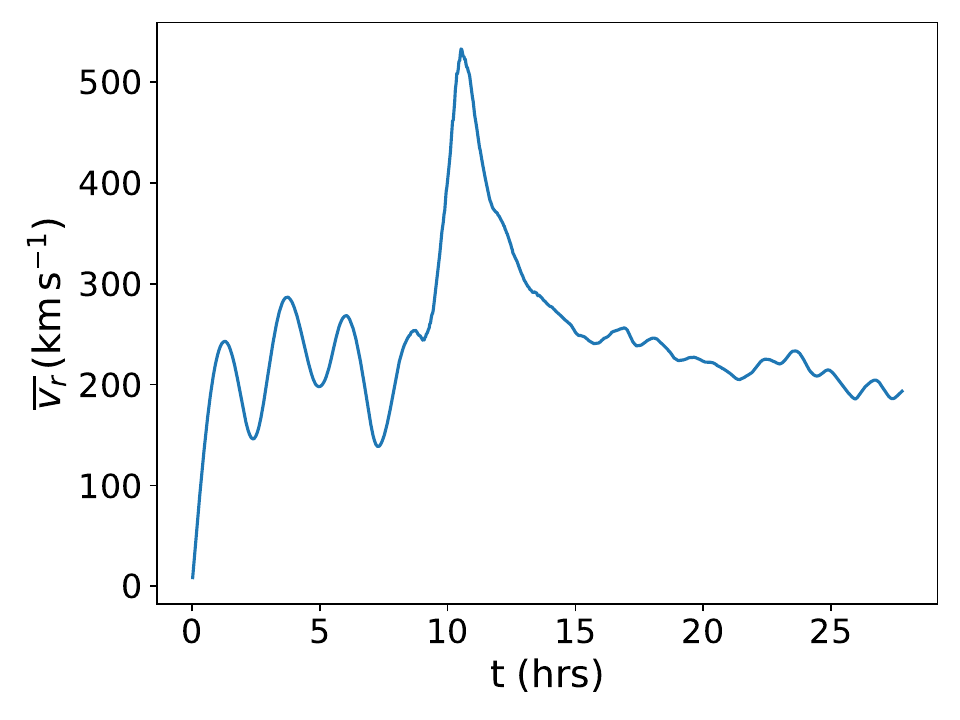}
    \caption{\label{fig:vavg_r} Average radial gas velocity as a function of radius at $t=14.17$ hours (top plot) and as a function of time around $4R_\odot$ (bottom plot).  As seen in the top plot, the radial velocity increases roughly linearly with radius before it asymptotes toward a value of $\overline{v_r} \sim 300\,{\rm km s^{-1}}$, which occurs roughly at $R=4R_\odot$.  The lower plot then observes the average radial velocity over time at this radius.  We see $\overline{v_r}$ fluctuate from $200-300\,{\rm km s^{-1}}$ before peaking at $\overline{v_r} \sim 500\,{\rm km s^{-1}}$ at $t\sim11{\rm hrs}$.}
    \end{figure}

\subsection{Angular Distribution of ejected mass}

\begin{figure}
\includegraphics[width=0.45\textwidth]{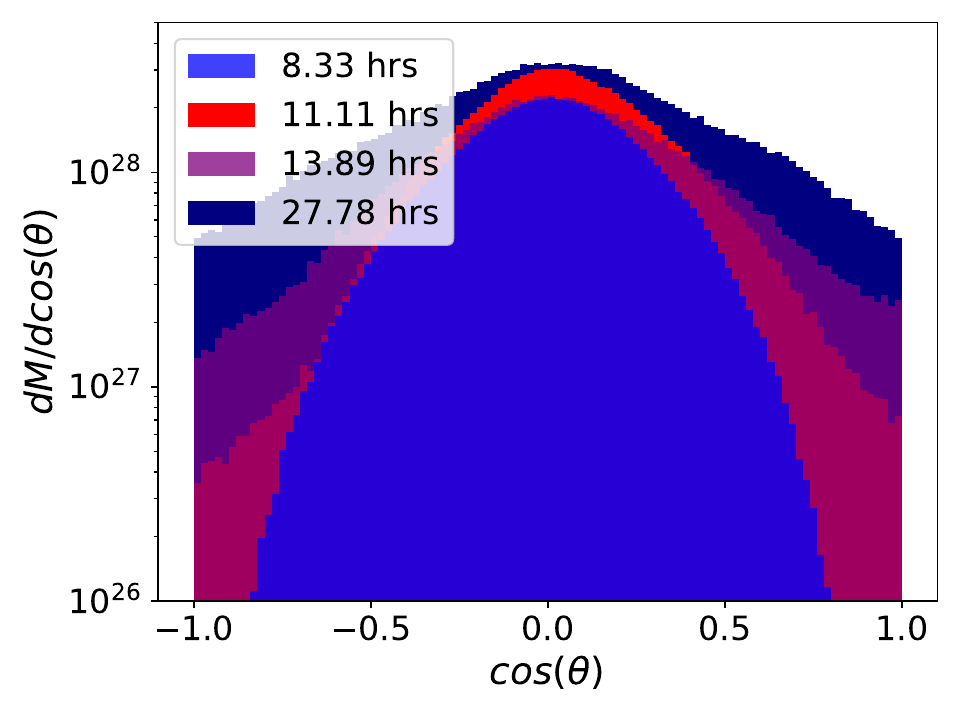}
\caption{\label{fig:multi_azi} Distribution of gas mass as a function of $\theta$ for $R>R_{Roche}$ at $t=8.33$, $11.11$, $13.89$, \textbf{ and $27.78$} hours.  The mass peaks in the orbital plane ($\cos\theta=0$) and is a minimum near the poles of the system ($\cos\theta=\pm1$).  The difference between the maximum and minima decreases as time increases but remains substantial at small radii.} 
\end{figure}

We now explore the angular distribution of the ejected material. In figure \ref{fig:multi_azi}, we plot the angular distribution of the mass, $dM/d\cos\theta$, for the material at $R>R_{\rm Roche}$ for $t=8.33$, $11.11$, $13.89$, and $27.78$ hours.  \textit{Because the material expands hydrostatically at these times, it} \textbf{ The material } flows through the ${\rm L}_1$ point\textit{, thus adopting} \textbf{ and adopts} a distribution that is concentrated near the orbital plane.  This is clearly seen in the peak of the distribution near $\theta=0$ and its deficit near the poles.  \textit{This is independent of time, as expected for a hydrostatically expanding envelope that overloads its Roche lobe at the ${\rm L}_1$ Lagrange point.}

The concentration near the orbital plane stands in stark contrast to the situation in common envelope evolution (CEE), where the ejecta is distributed over $4\pi$ steradians \cite{2023MNRAS.526.5365V}.  Part of it may be due to the difference in physical situation.  In the CEE situation, the secondary plunges through an envelope that is locally homogeneous in the plunging phase \cite{2019MNRAS.486.5809P,2023MNRAS.526.5365V}.  However, here the fluid is clearly not homogeneous as it pours over to the secondary at the ${\rm L}_1$ point.  It could also be a question of scale, as CEE ejection is roughly spherically at large radii.

\begin{figure}
    \includegraphics[width=0.45\textwidth]{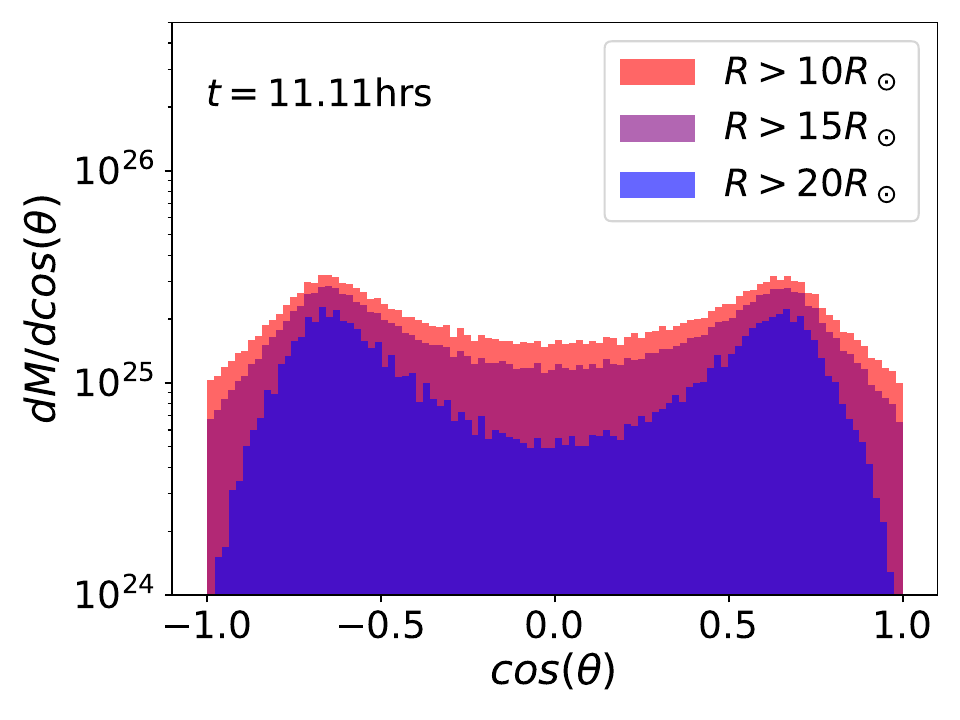}
    \includegraphics[width=0.45\textwidth]{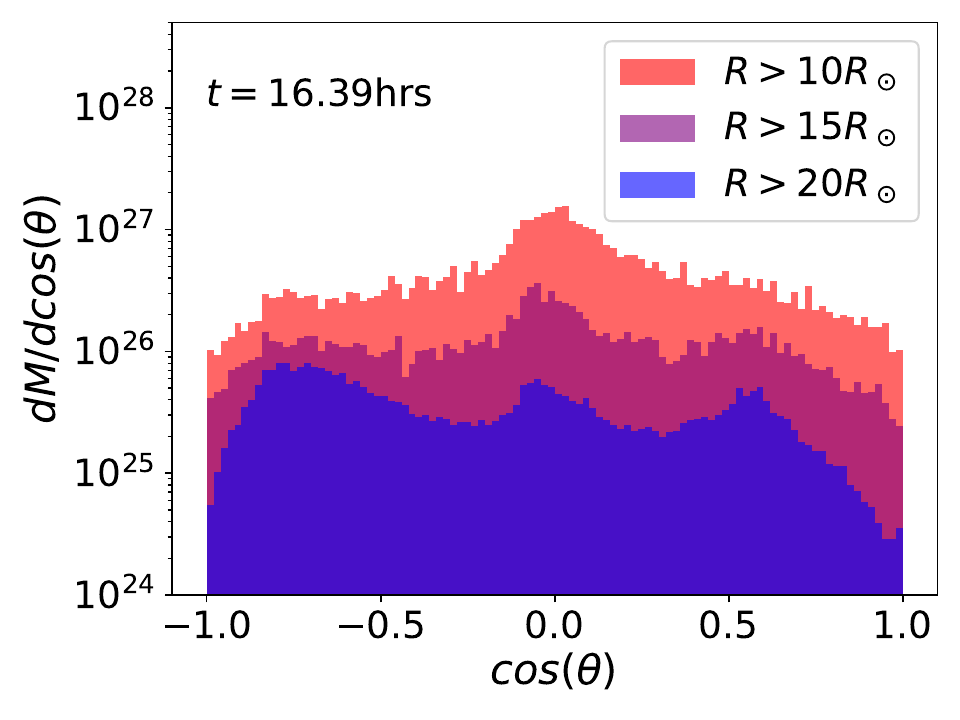}
    \includegraphics[width=0.45\textwidth]{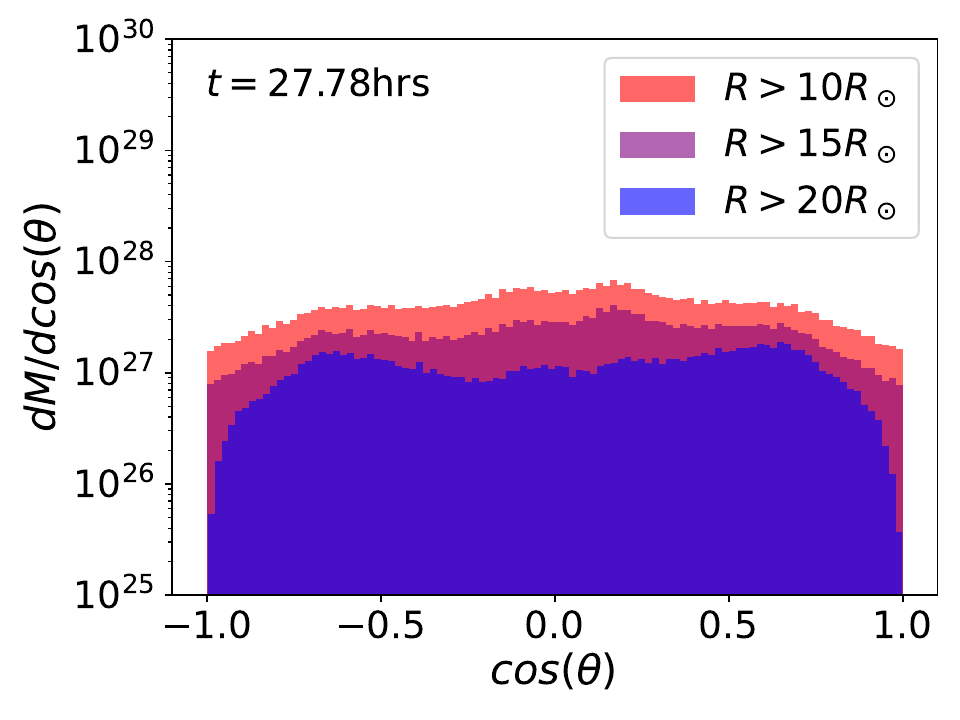}
    \caption{\label{fig:multi_rad_azi} Gas mass distribution for $R>10 R_\odot$, $R>15 R_\odot$, and $R>20 R_\odot$ \textbf{ at $t=11.11$, $16.39$, and $27.78$ hours.  For $t=11.11 \,{\rm hours}$, we see a deficit in material at the orbital plane ($\cos\theta=0$) with peaks between the plane and the poles ($\cos\theta=\pm1$) for all radii shown.}  For $t=16.39 \,{\rm hours}$, at $R>10 R_\odot$, the gas is slightly preferentially located toward the orbital plane by about one order of magnitude.  However, as the radius increases to $R>15 R_\odot$, we see this peak become only slightly greater than at the poles.  This is again seen at $R>20 R_\odot$, where the mass distribution is roughly spherical.  \textbf{ This mass distribution proves to be even more spherical at $t=27.78 \,{\rm hours}$ for all radii shown.}}
\end{figure}

There is indeed substantial evidence that at large radii the ejecta from classical novae is also roughly spherical.  In Figure \ref{fig:multi_rad_azi}, we plot the angular distribution of the mass for $R>10 R_{\odot}$, $R>15 R_{\odot}$, and $R>20 R_{\odot}$ at  $t=11.11$, $16.39$, \textbf{ and $27.78$ hours}.  At $t=16.39$ hours, for $R>10R_\odot$, the mass distribution is still biased toward the orbital plane, but this changes as we move toward $R > 20 R_\odot$, where the mass distribution is more isotropic.  Let us now compare this to the photosphere for these classical novae.  Given a mass loss rate of $0.01 M_{\odot}{\rm yr}^{-1}$, we can compute the density profile of the envelope from $\dot{M} = 4\pi \rho r^2v$ to find
\begin{equation}
\rho(r) = 10^{-6} \dot{M}_{-2}v_3^{-1}
r_{\odot}^{-2}
\end{equation}
where $\dot{M}_{-2} = {\dot{M}}/{10^{-2}M_{\odot}\,{\rm yr}^{-1}}$, $v_3 = {v}/{10^3 {\rm km\,s}^{-1}}$, and $r_{\odot} = {r}/R_{\odot}$.
The photosphere can be estimated to be where the optical depth $\tau = \kappa\rho r$, equals unity and thus:
\begin{equation}
r_{\rm ph} \approx 1.4\times 10^3 \dot{M}_{-2}v_{3}^{-1}\kappa_{0.2} R_{\odot},
\end{equation}
where $\kappa_{0.2} = {\kappa}/{0.2\, {\rm cm^2\,g^{-1}}}$.
This is roughly in line with observations \citep{2021ARA&A..59..391C}.  Notably, the photosphere is substantially outside of where the flow becomes spherical, and thus we can presume that the spherical distribution we see at 20 solar radii continues on to the photosphere.  \textbf{ Additionally, given ejecta velocities of around $v=200$-$500 \,{\rm km\,s^{-1}}$, and the $5$ hour time difference between the top and middle plot of Figure \ref{fig:multi_rad_azi}, the ejecta would flow $\sim5$-$13\,{\rm R_\odot}$.  This means that $R>10\,{\rm R_\odot}$ in the top plot will become $R>15\,{\rm R_\odot}$ or $R>20\,{\rm R_\odot}$ in the middle plot.  It is clear that this material becomes more spherical, indicating an evolution of the ejecta.  Even so, looking at $R>10\,{\rm R_\odot}$ in all three plots, the ejecta becomes more spherical as time progresses.  Thus, both effects of ejection and ejecta isotropization are present.}


\subsection{Angular Momentum }

\begin{figure}
    \includegraphics[width=0.45\textwidth]{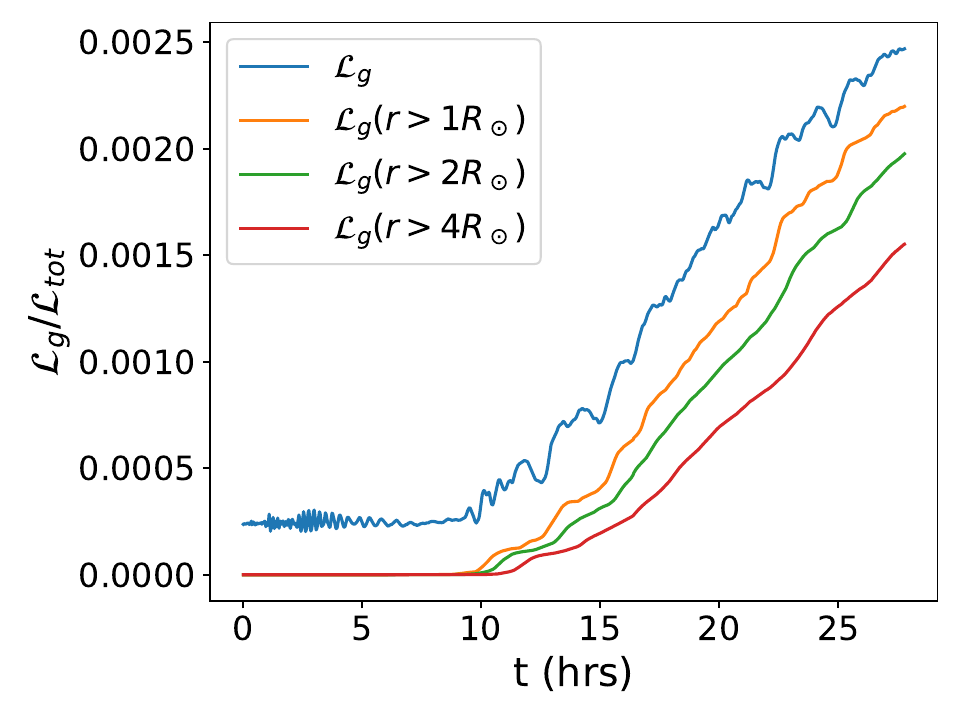}
    \caption{\label{fig:gas_L} Gas angular momentum as a function of time.  The blue line shows the total gas angular momentum, while the orange, green, and red lines show only the angular momentum of the gas beyond $1R_\odot$, $2R_\odot$, and $4R_\odot$ respectively.  The angular momentum of the gas remains roughly constant until ejection begins at $t\sim8.5\,{\rm hrs}$. The increase in the gas' angular momentum at this time is matched by a decrease in the binary's angular momentum.}
    \end{figure}
    
The ejection of mass during a nova carries angular momentum away from the binary, which can influence the evolution of the system \citep{2024ApJ...977...34T, 2011ApJS..194...28K, 2016MNRAS.455L..16S, 1991A&A...246...84L}.  We see this in Figure \ref{fig:gas_L} where we compute the angular momentum of the gas $\angmom_g$ relative to the total angular momentum of the \textbf{binary, where the angular momentum is defined relative to the center of mass of the binary}:
\begin{equation}\label{eq:total ang momentum}
    \angmom_{\rm tot}=\mu\sqrt{GM_{\rm tot}a}
\end{equation}
where $\mu=m_1m_2/M_{\rm tot}$ is the reduced mass and $M_{\rm tot}=m_1+m_2$ is the total mass, $m_1$ is the mass of the WD primary, and $m_2$ is the mass of the secondary. Note that $\angmom_g$ is initially nonzero. This is because the envelope is associated with the WD primary, which is orbiting about the center of mass of the binary.  However, $\angmom_g$ at $r>1$, $>2$, and $>4\,R_{\odot}$ starts at $0$ because the envelope does not reach these radii initially.  As the system evolves and the envelope expands and is ejected, $\angmom_g$ increases.  This must be matched by a corresponding decrease in the angular momentum of the binary.  

The magnitude of this effect depends on where the moment arm is.  The moment arm might be at the position of the WD, the secondary star, or the ${\rm L}_2$ Lagrange point.  To illustrate this, we first consider equation (\ref{eq:total ang momentum}) and expand around the secondary mass, $m_2\rightarrow m_2-\delta m$ to find:
\begin{equation}
    \delta \angmom = m_1(m_2-\delta m)\sqrt{\frac{Ga}{m_1+m_2-\delta m}}-\angmom_{\rm tot}.
\end{equation}
We then Taylor expand to find:
\begin{equation}
    \left.\frac{d\angmom}{dm}\right|_2=\frac{\angmom_{\rm tot}}{M_{\rm tot}}\left(\frac{1}{2}-\frac{1}{q_2}\right)
\end{equation}
where $q_2=m_2/M_{\rm tot}$.  A similar expansion around $m_1$ will yield an analogous expression for $\left.d\angmom/dm\right|_1$, except we replace $q_2$ with $q_1=m_1/M_{\rm tot}$.  

The expressions for $\left.d\angmom/dm\right|_1$ and $\left.d\angmom/dm\right|_2$ define the specific angular momentum loss via mass ejection around the primary and secondary, respectively. \citet{2024ApJ...977...34T} suggest that angular momentum can also be lost through the ${\rm L}_2$ point of the binary system.  The moment arm in this case is $r_{{\rm L}_2}$ which is the distance between the center of mass of the binary and the ${\rm L}_2$ point.  We express this as
\begin{equation}
    r_{{\rm com},{\rm L}_2}=r_{{\rm L}_2}-a+a(1-q_2),
\end{equation}
where $r_{{\rm L}_2}$ is the distance from the WD to the ${\rm L}_2$ point, and $a$ is the semi-major axis.  We then calculate the specific angular momentum lost through the ${\rm L}_2$ point with the following:
\begin{equation}
    \left.\frac{d\angmom}{dm}\right|_{{\rm L}_2}=\Omega r_{{\rm com},{\rm L}_2}^2,
\end{equation}
where $\Omega = \sqrt{GM_{\rm tot}/a^3}$ is the Keplerian frequency. 


\begin{figure}
    \includegraphics[width=0.45\textwidth]{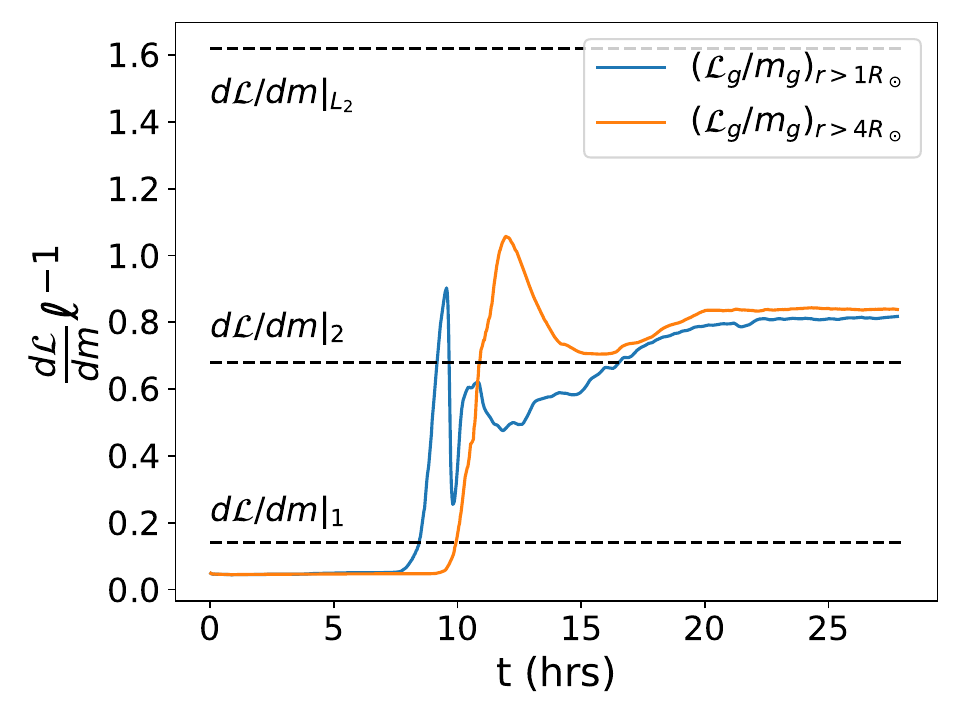}
    \caption{\label{fig:gas_specific_L} $d\angmom/dm$ normalized to \textbf{ $\ell$} as a function of time.  We measure the specific angular momentum of the gas beyond $1R_\odot$ and $4R_\odot$ and include predicted specific angular momentum at the position of the WD ($d\angmom/dm|_{1}$), the secondary ($d\angmom/dm|_{2}$), and the $\text{L}_2$ Lagrange point ($d\angmom/dm|_{L_2}$).}
    \end{figure}
    
In Figure \ref{fig:gas_specific_L}, we plot the specific angular momentum of the gas $\angmom_g/m_g$ for $r>1$ and $4R_{\odot}$ normalized to the specific angular momentum of the system,
\begin{equation}
    \ell = \sqrt{GM_{\rm tot}a},
\end{equation}
as a function of time.  Here, the specific angular momentum of the gas is $0$ until the gas flows across the $r>1$ and $>4R_{\odot}$ spherical surfaces at $t\approx 7.5$ and $10$ hours, respectively.  We also include the specific angular momentum at the primary $\left.d\angmom/dm\right|_1$, the secondary $\left.d\angmom/dm\right|_2$, and the ${\rm L}_2$ Lagrange point $\left.d\angmom/dm\right|_{{\rm L}_2}$.  The observed specific angular momentum loss is well predicted by $\left.d\angmom/dm\right|_2$, so we interpret the ejection of gas as an isotropic ejection at the position of the secondary.  The increase in the angular momentum of the gas would argue that the moment arm is not at the position of the WD primary.  Finally, we do not see angular momentum being lost through the ${\rm L}_2$ point of our system, which further reinforces our claim that the ${\rm L}_2$ Lagrange point is not important in nova ejection.

We will note that the angular momentum loss is modified by the binary interaction.  This is in qualitative agreement with the results of \citet{2022ApJ...938...31S} and \citet{2024ApJ...977...34T}.  However, the assumption that the mass is lost through the ${\rm L}_2$ point results in an enhanced angular momentum loss rate that is not supported by our simulations.  How this quantitatively affects the results of \citet{2024ApJ...977...34T} is a question beyond the scope of the current study.




\section{Discussion and Conclusions}\label{sec:discussion}

In this paper, we numerically explore the ejection of material by binary interaction in classical novae as recently explored by \citet{2022ApJ...938...31S}.  For a prescribed heating rate, we find that the binary interaction plays an important role in the expansion of the WD envelope.  Here we will summarize the salient features.  First, the envelope flows through the ${\rm L}_1$ Lagrange point and interacts with the secondary.  This interaction with the secondary drives the flow to be ejected from the system though it is ejected somewhat isotropically around the secondary, i.e. the ${\rm L}_2$ Lagrange point does not play an important role in the ejection.  We can see this in looking at the relative dominance of kinetic and thermal energies compared to the gravitational potential energies at the ${\rm L}_1$ and ${\rm L}_2$ Lagrange points.  Second, while the material is heavily influenced by the secondary and one might expect that a signature of the orbital plane would manifest itself in the ejecta distribution, we find that at large radii the distribution is more or less spherical.  This is in line with recent numerical studies on CEE, which found a spherical distribution of ejecta at large distances \cite{2023MNRAS.526.5365V}.  Finally, the interaction with the binary enhances the amount of angular momentum carried away by the ejecta as it changes the moment arm to the location of the low-mass secondary.  This will likely enhance the angular momentum transport and evolution of binary systems, but the effect would be moderated compared to the results of \citet{2024ApJ...977...34T}.

There are a number of future directions in which this can be taken.  First, since the ejection is expected to be spherical at large distances, a study of how this ejected material continues to evolve would be a worthy topic of study.  In particular, the spherical distribution suggests that a simple 1-D study may be sufficient to capture all the dynamics.  Second, how this enhanced angular momentum transport affects the evolution of the binary system is also a worthwhile topic of study. Finally, we have assumed that the secondary is essentially a point-like gravitational mass in this work.  In reality, the secondary is a large Roche-filling body. The hydrodynamic interaction between the expanding WD envelope and the Roche-filling secondary cannot be discounted in this situation.  The effect of hydrodynamic effects on the expanding envelope is a challenging numerical problem but is a topic for future study.

\begin{acknowledgments}
PC thanks K. Shen for giving a really nice talk that inspired this work.
NN acknowledges the support of the Northwestern Mutual Data Science Institute through the Student Scholar award and the University of Wisconsin-Milwaukee through the Support for Undergraduate Research Fellows and Senior Excellence in Research Award. PC acknowledges support from the National Science Foundation through awards AST-2307885 and AST-2108269 and from the NASA ATP program through award 80NSSC24K0936.  We also acknowledge support from the CIberCATSS program supported from the NSF via award OAC-2229652. \changaMM 
simulations were completed on the Mortimer HPC System at UWM, which was funded in part by the
NSF Campus Cyberinfrastructure Award OAC-2126229 and UWM.
\end{acknowledgments}

\begin{center}
\textbf{DATA AVAILABILITY STATEMENT}
\end{center}
The code required to reproduce the result of this paper is available at this github repository: \url{https://github.com/N-BodyShop/changa}.  A compiled ready-to-run version is available at \url{https://www.cambercloud.com}.  Please contact the authors for setup instructions.

The simulation data that support the findings of this article are not publicly available, but are available from the authors upon reasonable request.

\nocite{*} 
\bibliography{references} 
\end{document}